\documentclass[jgr,pdftex]{agu2001}
	\usepackage[dvips]{graphicx}            
	\usepackage{pslatex}	    
	
\authorrunninghead{OUILLON AND SORNETTE}

\titlerunninghead{FAULT RECONSTRUCTION}

\authoraddr{Guy Ouillon, Lithophyse, 4 rue de l'Ancien S\'enat, 06300 Nice, France 
(e-mail: lithophyse@neuf.fr)}
\authoraddr{Didier Sornette, D-MTEC, and Department of Earth Sciences, ETH Z\"urich, Kreuzplatz 5,
CH-8032 Z\"urich, Switzerland and D-ESS and IGPP, UCLA, Los Angeles, California, USA
(e-mail: dsornette@ethz.ch)}

\begin{document}

\title{Segmentation of Fault Networks Determined from Spatial Clustering of Earthquakes}

\authors{G. Ouillon, \altaffilmark{1} and D. Sornette\altaffilmark{2,3}}

\altaffiltext{1}{Lithophyse, 4 rue de l'Ancien S\'enat, 06300 Nice, France}

\altaffiltext{2}{D-MTEC, and Department of Earth Sciences,
ETH Z\"urich, Kreuzplatz 5, CH-8032 Z\"urich, Switzerland}

\altaffiltext{3}{Department of Earth and Space Sciences and
Institute of Geophysics and Planetary Physics, 
University of California, Los Angeles, California 90095-1567, USA}

\newcommand{\be}{\begin{equation}}
\newcommand{\ee}{\end{equation}}
\newcommand{\ba}{\begin{eqnarray}}
\newcommand{\ea}{\end{eqnarray}}
\newenvironment{technical}{\begin{quotation}\small}{\end{quotation}}
\renewcommand{\thefootnote}{\arabic{footnote}}
\newtheorem{definition}{Hypothesis}


\begin{abstract}
We present a new method of data clustering applied to earthquake catalogs, with the goal
of reconstructing the seismically active part of fault networks. 
We first use an original method to separate clustered events from uncorrelated seismicity using
the distribution of volumes of tetrahedra defined by closest neighbor events in the original and 
randomized seismic catalogs. 
The spatial disorder of the complex geometry of fault networks is then taken into account
by defining faults as probabilistic anisotropic kernels, whose structures are motivated
by properties of discontinuous tectonic deformation and previous empirical observations of the 
geometry of faults and of earthquake clusters at many spatial and temporal scales. 
Combining this a priori knowledge with information theoretical arguments, 
we propose the Gaussian mixture approach implemented in an Expectation-Maximization (EM) procedure. 
A cross-validation scheme is then used and allows the determination of the number of kernels that should be used
to provide an optimal data clustering of the catalog.
This three-steps approach is applied to a high quality relocated catalog of the seismicity following the 1986 Mount Lewis ($M_l=5.7$) event in California
and reveals that events cluster along planar patches of about $2$~km$^2$, i.e. comparable to the size of the main event. The finite thickness of those clusters
(about $290$~m) suggests that events do not occur on well-defined euclidean fault core surfaces, but rather that the damage
zone surrounding faults may be seismically active at depth. Finally, we propose a connection between our methodology 
and multi-scale spatial analysis, based on the derivation of  spatial fractal dimension of about $1.8$ for the set of hypocenters in the Mnt Lewis area, consistent with recent observations on relocated catalogs.  
\end{abstract}


\begin{article}

\section{Introduction \label{sintro}}

Earthquakes are thought to be clustered over a very wide spectrum of spatial scales 
(see Kagan and Knopoff (1978; 1980), Kagan (1981a; 1981b; 1994) Sornette (1991; 2006), Sornette et al (1990),  for the main concepts and observations). At the largest, worldwide scales, the dominating feature of earthquake epicenters maps is the existence of narrow corridors of activity, coinciding with tectonic plates boundaries. Observation of earthquake distribution at smaller and smaller scales (down to the standard accuracy of the spatial locations, typically a few kilometers - or below in the case of relative locations) successively reveals smaller-scale corridors (or surfaces in 3D), qualified as faults by geologists and seismologists.

The identification of faults and of the statistical properties of their associated seismicity is a major challenge in order to delineate the potential sources of future large events that may have a societal impact. It is also necessary to clearly understand the mechanical interactions within a given fault network, in order to possibly achieve earthquake prediction with a reasonable accuracy, both in the space and time dimensions. On more fundamental grounds, understanding the relationships between faults (or more generally earthquake generating cluster centers) at different scales can help to 
develop a meta-theory based on the renormalization group approach (Wilson, 1975 ; All\`egre et al, 1982) to progress on the understanding and modeling of seismicity and rupture processes as a critical phenomenon (Sornette, 2006; Bowman et al, 1998 ; Weatherley et al, 2003), as the size of an upcoming event may be correlated with the range of scales over which earthquake clustering properties change collectively (thus yielding a measure of the correlation length).

Fault identification from seismicity alone is a complex problem, whose success mainly depends on the seismotectonic experience and intuition of the operator, as it demands to incorporate data possessing dimensions of different natures (spatial location of hypocenters, rupture lengths or surfaces, magnitudes or seismic moments, and focal mechanisms of earthquakes), some of which being determined with variable and sometimes large uncertainties or ambiguity (like focal mechanism, for instance). Some properties of seismicity being highly heterogeneous even on well-identified faults, the identification thus essentially remains a human process, without automatization (which would require quantitative criteria to be fulfilled).

The present paper will focus on the automatic identification of faults from the sole locations of seismic events, as
recorded in a standard earthquake catalog, assuming that earthquake clusters reveal the underlying existence of such tectonic structures. This hypothesis is now universally accepted, at least for shallow events. We shall thus first provide the main definitions of clustering that will be used all along this work in section \ref{ssdefclust}, as those definitions may differ in the pattern recognition and the geosciences fields respectively. We refer to (Ouillon et al, 2008) for 
a review of existing data clustering techniques applied to earthquakes. 
We advise the reader to go to the related section in (Ouillon et al, 2008) 
as it underpins the argumentation developed in the present article. Section \ref{ssconj} will then present some very general physical and morphological arguments about the nature of the tectonic strain field, which will help us choosing the appropriate data clustering methodology. This method will be presented in section \ref{smixture}. It necessitates the introduction of a probabilistic kernel that defines the spatial structure of the set of events triggered by a given fault. Section \ref{ssclustfaultzone} will present the theoretical and empirical knowledge about earthquakes and faults in order to choose more rigorously the explicit shape of this probabilistic kernel. Section \ref{sgaussmix} will use the results presented in the previous section together with some information theoretical arguments in order to provide an analytical expression for the probabilistic kernel. The potential flaws of this kernel will be outlined and discussed as well.  The resulting data clustering procedure will then be applied to a synthetic example in section \ref{appsynthex} where we also introduce a cross-validation scheme that allows us to determine objectively the total number of kernels that should be used to fit the catalog. Section \ref{appmntlew} focuses on the natural catalog of the Mount Lewis area, following the $M5.7$ event that occurred in $1986$. Its analysis requires a preliminary treatment to remove events that occurred on faults that are under-sampled by seismicity. The application of our three-steps methodology to this natural catalog provides important informations about the possible structure of fault zone segmentation and activity at depth that are discussed in section \ref{conclusion}.



\section{Standard clustering techniques \label{ssdefclust}}

\subsection{What is clustering?}

It is useful to make precise the definition of the term ``clustering'' that will be used throughout this paper.
Indeed, while they study the spatial and/or temporal properties of the same set of events, 
seismologists and statisticians may not use the term ``clustering'' exactly in the same sense. 

In statistical seismology, {\it clustering} refers to the concentration of earthquakes into groups that form 
locally dependent structures in time and space.
The most obvious form of space and time clustering is the pattern defined by aftershocks.
Identifying the clusters then often amounts to determining the so-called main events,
which are interpreted as the triggers of the following aftershocks which aggregate
around the former in time and space. This procedure consisting in identifying
the main shocks is indeed usually defined as {\it declustering} in the seismological literature,
i.e., the reduction of a whole catalog to the set of the main shocks by removing the
corresponding aftershocks. There are several standard declustering methods
(Gardner and Knopoff, 1974; Knopoff et al., 1982; Reasenberg, 1985; Zhuang et al., 2002;2004; 
Marsan and Lenglin\'e, 2008; Pisarenko et al., 2008), but recent developments suggests
that the distinction between main shocks and aftershocks may be more a
human construction than reflecting a genuine property of the system (Helmstetter and Sornette, 2003a; b). 
In any case, earthquake clustering considers that correlations
between earthquakes stem from causal relationships between events.

In contrast, statistical pattern recognition aims at determining the local similarities or correlations between events 
in a given data set (Bishop, 2007; Duda et al., 2001). Those similarities may have different natures, such as size, color, spatial location and so on. The identification of clusters then consists in grouping together events that share those similarities, without choosing any specific event that would be genetically responsible for all the others. This process is called {\it clustering}. Clearly, it considers that correlations among events stem from a common cause, i.e. the existence of a common underlying fault. 

As those terminologies can be rather confusing, we shall use the terms {\it earthquake clustering} when dealing with the phenomenology of seismic events triggering other seismic events, and {\it data clustering} when dealing with the identification of clusters. Our focus is to determine earthquakes that are clustered in the second sense, with each cluster being a candidate fault, i.e., a cluster of events will be conjectured to occur on the same fault.
In contrast, the events involved in earthquake declustering may occur on different faults.

\subsection{Main categories of data clustering techniques}

There are many available pattern recognition techniques for data clustering. Rather than
presenting an exhaustive review, the interested reader is referred to
Bishop (2007) and Duda et al. (2001), which describe a large number of 
data clustering techniques. Some others, which have been applied to earthquakes, are briefly discussed by
Ouillon et al. (2008). From now on, we shall focus exclusively on data clustering techniques 
in the spatial domain, except when explicitly mentioned.

Ouillon et al. (2008) have suggested a clustering algorithm that fits an earthquake catalog with 
rectangular plane segments of adjustable sizes and orientations. This method
is a generalization to anisotropic sets of the popular $k$-means method (McQueen, 1967).
Ouillon et al. (2008) proposed an iterative version, 
in the sense that planes are introduced one by one in the fitting procedure until the discrepancy between the proposed fault pattern and the spatial structure of the earthquake catalog can be explained by location errors only. Ouillon et al. (2008) found that the final fault pattern provided by the fit is in general in rather good
agreement with the fault pattern inferred by other methods, when available (such as maps of surface fault traces). However, the assumption that
fault elements are perfect rectangular plane segments is certainly an oversimplification. One of the
advantage of the method presented here is to improve on this reductionist assumption.

The main drawback shared by previous clustering approaches, and also but to a 
lesser degree by the anisotropic $k$-means of 
Ouillon et al. (2008), is that these methods are indiscriminately applied to earthquake catalogs
without recognizing their specific characteristics. As a consequence, it is difficult to decide
a priori which cluster solutions are to be chosen among the various constructions offered
by the different clustering techniques. In this paper, we propose to incorporate the existing knowledge about earthquakes and faults into the definition of the data clustering technique to be applied to earthquake datasets. 
The next section presents a first general argumentation based on very general and qualitative physical and geological 
properties, while more empirical and quantitative issues will be discussed in section \ref{ssclustfaultzone}. The available knowledge as well as the identification of the unknown will be 
combined with arguments from information theory to form the basis of our approach.

At its most basic level, data clustering consists in identifying `unusual' discrepancies 
between a spatial distribution of data and a smooth reference distribution. The homogeneous Poisson distribution is often considered as the reference distribution, so that clusters emerge as anomalies from a statistically uniform background. This idea is very intuitive but still remains to be proven as adequate for earthquakes. We shall discuss in a later section the statistical properties of the reference distribution we should consider for the case of earthquakes, using simple arguments about the symmetry of the involved physics and of the rheological nature of the Earth's crust.


\section{Some preliminary conjectures about the probable nature of earthquake clusters \label{ssconj}}

We propose that an appropriate data clustering procedure for earthquakes should be informed by the probable morphological nature of individual earthquakes as well as their relationship with the strain field within a tectonic plate. The time-varying strain field imposed at the boundaries of a given tectonic domain is partially accommodated within this domain through a variety of continuous rheological processes. Elastic, viscoelastic and viscoplastic deformation mechanisms provide a smooth (i.e. regular) spatial variation of strain that can, for instance, be monitored using techniques such as GPS, leveling, strain rosettes, and so on. Such data are generally sampled at a rather coarse resolution and concern only very recent periods of time, especially for the most sophisticated and accurate ones. The regularity of their spatial variation ensures that they can be interpolated with sufficient confidence both in space and time. 

In addition, rupture processes or surface friction instabilities contribute in general significantly to the 
overall strain balance. Compared with the continuous rheological processes, 
they correspond to mathematical singularities, labelled as {\it earthquakes} at small time-scales and {\it faults} at longer time-scales. One can typically observe only a very small number of such strain field discontinuities, as a result of instrumental limitations preventing their detection. For example, in any area, there exists a minimum magnitude below which one fails to detect earthquakes in a systematic manner. The same holds true for small faults 
or as a result of vegetal or urban masks at the surface. It is 
commonly believed that the observed singularities represent only a very small sample of a much larger set, that is, the issue of missing and censored data is acute.

Statistical seismology and tectonics have to deal continuously with drastically under-sampled and missing data. The impact of under sampling may be different, 
depending on the property to characterize. In the case of earthquakes, if we assume that their magnitude distribution obeys the Gutenberg-Richter law, this impact depends solely on the $b$-value. We shall hereafter assume that it is equal to its observed standard value, i.e. $b=1$. If one is interested in energy budgets or in strain fluctuations, the largest earthquakes dominate the balance. Therefore, in this case, 
the under sampling of small earthquakes has a relatively minor impact.
If we are interested in estimating the stress fluctuations within a given spatial domain, all magnitude ranges are believed to contribute with 
similar amplitude to the balance. Of course, there are subtleties in the
way the sizes of earthquakes control the spatial dependence of the 
stress fluctuations, as large events are probably associated 
with longer range stress correlation while small events exhibit
stronger collective spatial dispersion. In that case, undersampling affects the stress balance proportionnally to the amount of missed events.

As we are interested in the geometry of the set of event hypocenters, we have to
recognize that the smallest events dominate by far in the catalog 
as they are much more numerous. However, they are conspicuously
under sampled and actually absent below the magnitude detection threshold.
For example, assume that the Gutenberg-Richter law holds down to magnitude $M=-1$ and that the magnitude detection threshold is $M=2$ as is typical. Then,
the job of determining the spatial organization of earthquakes and 
the fault network on which they occur has to be carried out 
on just 0.1\% of the total number of events, while $99.9\%$ of the 
earthquakes are missing in the available catalog. This would have a minor impact on statistical analyses if all clusters featured a similar spatial density of events. However, seismicity is known to behave highly intermittently both in time and space, so that the characterization of regions of low activity is definitely penalized by the under-sampling.
This pessimistic observation must be balanced by the fact that the recorded seismic events reveal some of the most correlated part of the singular strain field, as they propagate over larger distances. These located events occur on a subset of the fault network
that is, by construction, associated with potentially dangerous earthquakes. In that sense, reconstructing the structure that supports such events certainly provides meaningful information for the quantitative estimation of seismic hazard. This conclusion is supported by Hauksson et al. (2006), who compared the locations of earthquakes and those of the known major fault in California.
They observed that, on average, larger events stand closer to the major faults than smaller events do.


\section{Decomposing a dataset as a mixture \label{smixture}}

\subsection{Complete vs. incomplete datasets \label{sspresprob}}

\begin{figure}
\noindent\includegraphics[width=15cm]{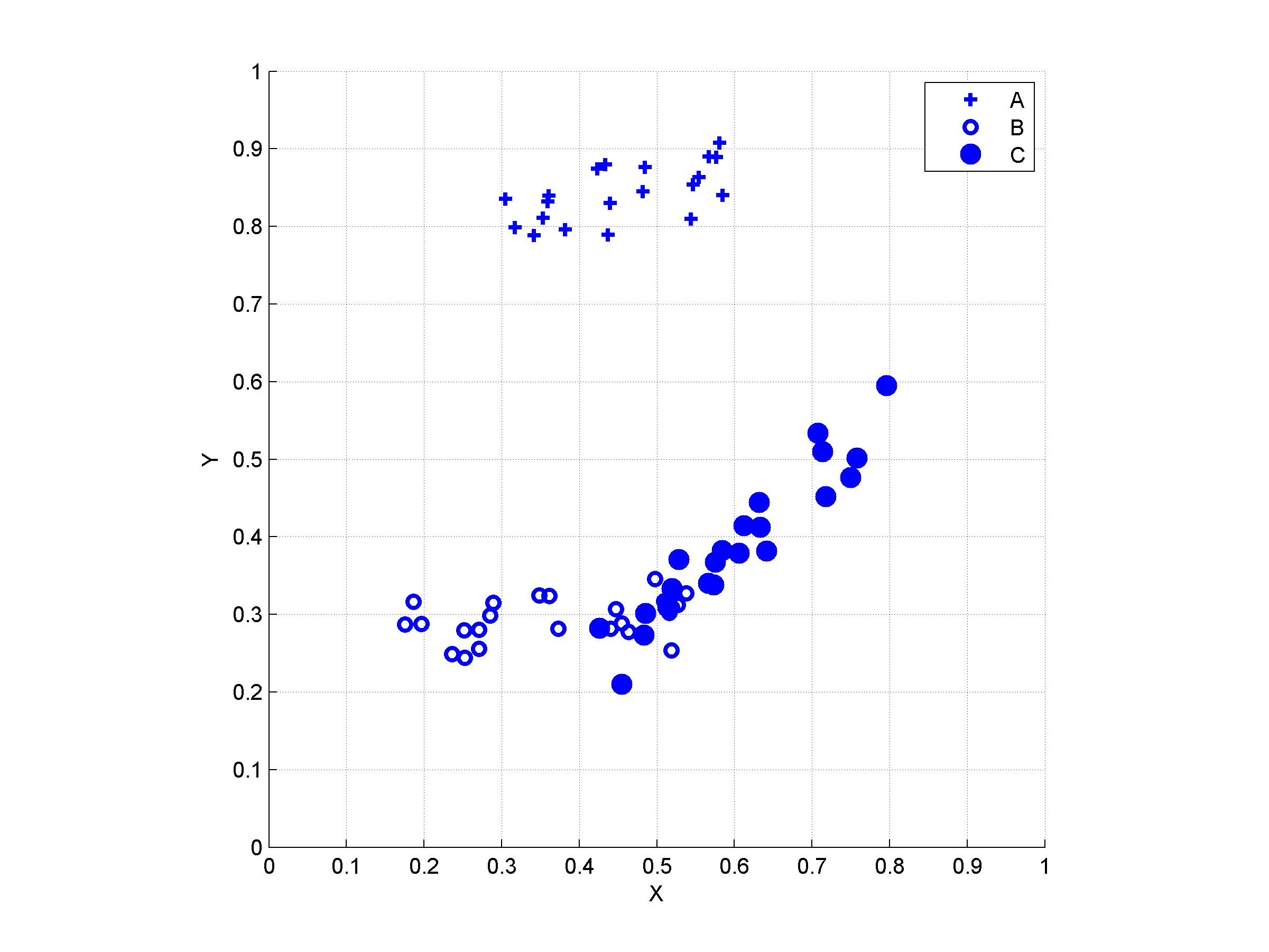}
\caption{\label{figclusta} Synthetic $2D$ dataset featuring $3$ clusters. Each event is represented by a symbol which allows to identify which cluster it belongs to ($A$, $B$ or $C$).}
\end{figure}

\begin{figure}
\noindent\includegraphics[width=15cm]{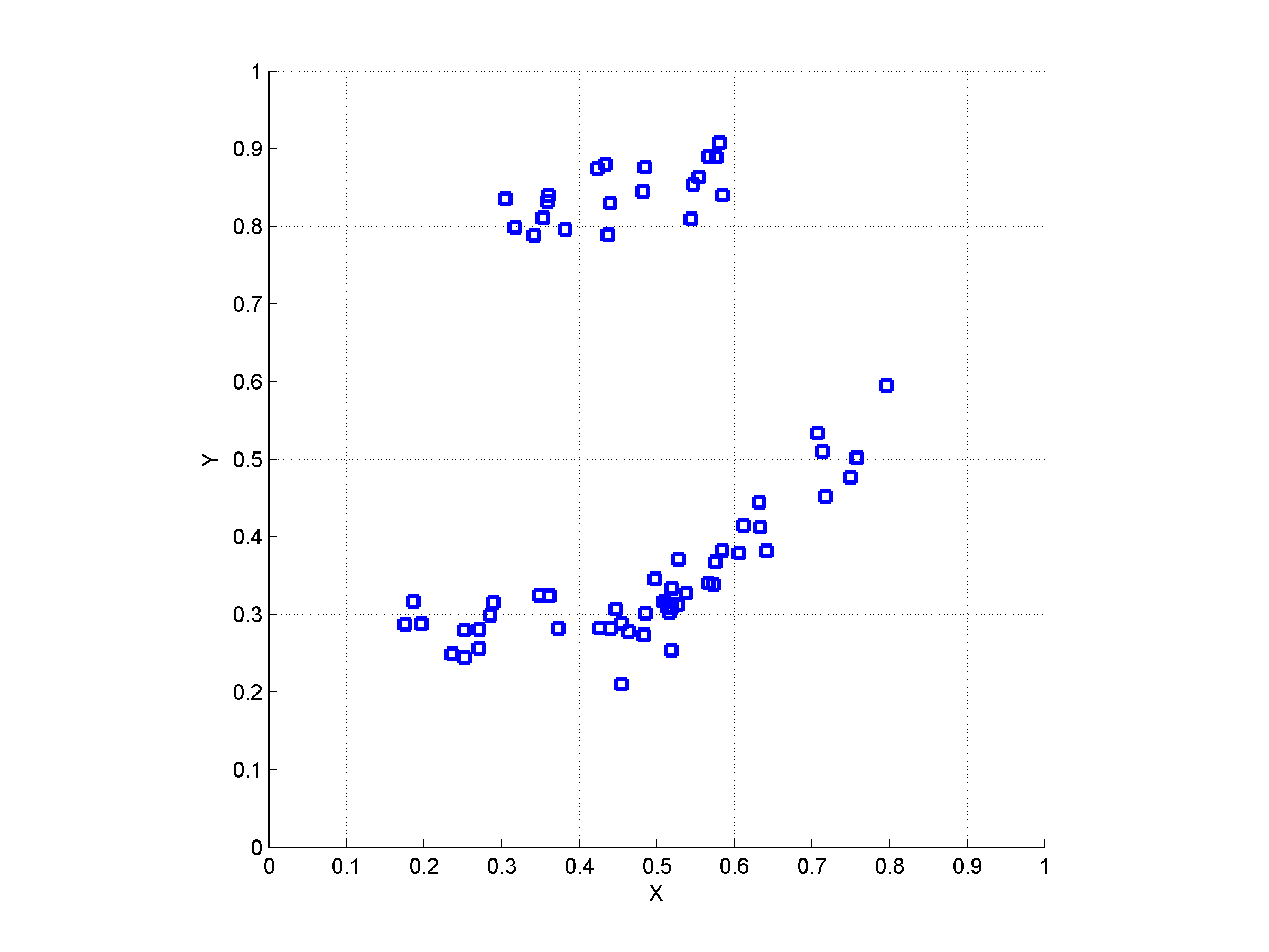}
\caption{\label{figclustb} Same dataset as in Fig. \ref{figclusta}, removing the genetic origin of events.}
\end{figure}

The generic problem we address is schematically represented 
by comparing Fig.~\ref{figclusta} with  Fig.~\ref{figclustb}.

In Fig.~\ref{figclusta}, three synthetic event clusters are represented
in a two-dimensional projection (corresponding to earthquake epicenters).
All the considerations and methods discussed apply to 
the 3D case without difficulty. The three synthetic clusters
represent model earthquakes occurring on three distinct faults.
The three clusters were generated as follows: (i) define three linear segments over which data points are randomly located, and (ii) add a random perturbation to the position of each point, yet keeping the memory of the linear segment over which the point was initially located. 

Each event is represented using a different symbol (empty circle, full circle or cross) according to which cluster (fault) it belongs too. Reciprocally, events with the same mark (i.e. represented by a common symbol) inform us that they should be classified within a common cluster. We thus know that all events represented by an empty circle belong to the same cluster (that we label $A$, for instance), all events plotted using a full circle belong to the same cluster $B$, while all events plotted with a cross belong to the same cluster $C$. Each event is thus characterized by a position $(x,y)$ and a label ($A$, $B$ or $C$), thus defining a {\it complete dataset} (see Bishop, 2007). Using this complete dataset, it is then very easy to provide estimates of the characteristics of the clusters $A$, $B$ and $C$, such as their spatial location, size and orientation. It is then also straightforward to compute for each cluster the statistical properties of any mark its events could carry. 

Earthquake catalogs do not belong to this category, in the sense that events that are recorded do not naturally feature any label informing us about the fault on which they occurred or that they occurred on a common fault. The typical data set of epicenters is exemplified by the set shown in Fig.~\ref{figclustb}. This is the same set as the one represented in Fig.~\ref{figclusta}. Because we do not have the luxury of knowing
a priori neither the generating fault nor any information on cluster association, all events are represented by a common symbol (a square), which informs us only on the position of each event. 
When going from Fig.~\ref{figclusta} to  Fig.~\ref{figclustb}, we have also lost the information on how many faults are present 
in the system. The event locations alone define the {\it incomplete dataset}, while the missing information (labels and number of labels) define the {\it latent variables} (see Bishop, 2007). Our task is to process the data provided in Fig.~\ref{figclustb} to attempt recovering the original partition of Fig.~\ref{figclusta}, {\it i.e.}, to explain the location of the events by estimating a solution for the latent variables. This problem is perhaps obvious for events that are plotted using crosses on Fig.~\ref{figclusta}, as they still appear as a clearly isolated cluster in Fig.~\ref{figclustb}. But even here, one could eye-ball two or three sub-clusters, suggesting the need for three faults for just this cluster. This gives a first taste of the role of location errors in the detection of relevant structures. The problem is even more difficult for the events shown as empty or full circles in Fig.\ref{figclusta}, as the two clusters overlap. 

In the following, we first work out the methodology to recover the clusters and faults, given the knowledge of their number ($3$ in the example of Fig. \ref{figclusta} and Fig. \ref{figclustb}). As this number is unknown in real seismic catalogs, we then develop a method to determine this value from the dataset, which is presented in section \ref{appsynthex}. This methodology will also take account of individual earthquake location uncertainities and will evaluate the significance of the proposed data clustering solution.

\subsection{Hard vs. soft assignment \label{sshardsoft}}

One can distinguish two broad classes of clustering approaches, called
respectively ``hard'' and ``soft'' assignment methods. Hard assignment characterizes the data clustering methods such as {\it K-means} (MacQueen, 1967). See Ouillon et al. (2008) for a general presentation of the method and its extension to the representation with planes of 
the spatial organization of earthquakes recorded in catalogs. {\it K-means} is an iterative and non-probabilistic method that partitions a set of $N$ datapoints into a subset of $K$ clusters, each cluster being characterized by its barycenter and variance (or its best fitting plane and covariance matrix as in Ouillon et al, 2008). {\it K-means} aims at minimizing the global variance of the partition, and provides at least locally reasonable, if not absolute, minima.
When convergence has been reached, each event belongs to the cluster defined by the closest barycenter (or plane) with probability $P=1$, while it has a zero probability of belonging to any other cluster. This way of partitioning data certainly helps in providing a rough approximation of the partition, but clearly lacks subtlety when clusters overlap or when location uncertainty is large. For instance, for the set 
of events shown in Fig.~\ref{figclustb}, in the region where the two clusters $A$ and $B$ intersect each other, 
the hard assignment method is bound to incorrectly attribute
with probability $P=1$ some events to cluster $A$ whereas they indeed belong to cluster $B$, and vice-versa. 

In contrast, soft assignment gives for each event $n$ the set of probabilities $P_n(A), P_n(B)$ and $P_n(C)$ that it belongs to $A$, $B$ or $C$, with $P_n(A)+P_n(B)+P_n(C)=1$. In that approach, each cluster consists in a more or less localized probabilistic kernel so that the full dataset is considered as a stochastic realization of a linear {\it mixture} of such probabilistic kernels. If the spatial domain over which the kernel is defined is infinite, then any point located at a large distance from a cluster has a non zero probability to belong to it. Certainty is thus banished from this view. It seems nevertheless well-adapted to those cases where data are located with uncertainty, as encountered in earthquake catalogs. We shall see later that there exist much deeper reasons for using a probabilistic approach to the data clustering of seismicity. An important notion underlying the mixture approach is that the events within each cluster are independent identically distributed.
Any pattern in the set of data points is assumed to be accounted for by the property of belonging to one of the set of clusters. Thus, within a cluster, 
all events are assumed to play the same role and are only related to each other through the property of belonging to this cluster. In other words, 
any spatial correlation among events featuring the same label is considered to be only due to the fact that they have been generated by a common localized kernel, so that events are correlated because they have the same cause. The idea of events triggering (or shadowing) some other events is thus completely foreign to this description. This notion indeed holds for the synthetic set of events shown on Fig.\ref{figclusta} as none of those events has been generated as a function of the position of the other events. 

In statistical seismology, this would mean that preexisting faults generate earthquakes, but that earthquakes do not interact with each other, independently of the fact that they belong to the same fault or to different faults. A proper use of this technique would thus require first to decluster (in the seismological sense) an earthquake catalog, thus identifying all triggered events, and to keep only the set of independent events to be decomposed as a mixture. Such declustering methods exist, as mentioned above, and are usually employed to study aftershock sequences. Unfortunately, they are either dependent on a set of arbitrary parameters (such as the size of spatial and temporal windows within which two events are considered as causally correlated, and their dependence on the size of the first, triggering event), or on a pre-assigned mathematical model (as in the ETAS model; see (Helmstetter and Sornette, 2002) and references therein). In the end, the result may also depend heavily on the magnitude detection threshold (Sornette and Werner, 2005b). The lower this threshold, the smaller the number of events that are qualified as independent, so that too few earthquakes could be used in a data clustering scheme.

\subsection{The expectation-maximization (EM) algorithm \label{ssem}}

The expectation-maximization (or EM) algorithm is an iterative method that allows one
to perform the mixture decomposition mentioned in the previous sub-section.  In its most standard form,
it requires several assumptions (Bishop, 2007 ; Duda et al, 2001):

\begin{itemize}
\item The number of kernels to be used to partition the data is known and set to $K$.

\item All kernels used to partition the data have the same (or similar) analytical shape(s). For example, all kernels are Gaussian, yet their mean vectors and covariance matrices can differ from cluster to cluster.

\item The resulting partition is a distribution which consists in the sum of all kernels, so that the decomposition is linear.

\end{itemize}

All those assumptions will be discussed later. It is rather easy to relax the two first ones while 
the last assumption is more crucial.
As we plan to determine the position and size of clusters, we shall assume that the mean vector $\vec{\mu}$ and covariance matrix $\Sigma$ of each kernel exist, and we shall refer to a given kernel as $F(\vec{x} | \vec{\mu}, \Sigma)$.
More generally, the kernel could be expressed as a function of higher-order statistical moment tensors. Here, $F$ is a generic kernel and is often chosen as a multivariate Gaussian with unit variance in every direction. We shall check in section \ref{sschoice} 
the degree to which this is a correct assumption for earthquakes and faults.

The EM algorithm assumes that, at each location $\vec{x}$, the local spatial probability density of events is given by:

\be
p(\vec{x}) = \sum_{k=1}^{K} \pi_k F(\vec{x} \mid \vec{\mu_k}, \Sigma_k)~,
\ee

where the {\it mixing coefficients} $\pi_k$ are such that

\be
0 \leq \pi_k \leq 1, ~~~~\forall k~,
\ee

together with

\be
\sum_{k=1}^{K} \pi_k = 1~.
\ee

The $\pi_k$'s measure the relative weight of each kernel in the mixture. The two last equations ensure that

\be
\int_{space} p(\vec{x}) d\vec{x} = 1~.
\ee

The next important quantity is the set of {\it responsibilities} defined as

\be
\gamma(k \rightarrow n) = \frac{\pi_k F(\vec{x} \mid \vec{\mu_k}, \Sigma_k)}{\sum_{j=1}^{K} \pi_j F(\vec{x} \mid \vec{\mu_j}, \Sigma_j)}~.
\label{resp}
\ee

Thus, $\gamma(k \rightarrow n)$ is the probability that event 
$n$ is explained by kernel $k$. $\gamma(k \rightarrow n)$ thus belongs to $[0;1]$. One can then also define

\be
N_k = \sum_{n=1}^{N} \gamma(k \rightarrow n)~,
\label{nk}
\ee

as the expected total number of events explained by kernel $k$. It is thus not necessary an integer. We obviously have:

\be
\sum_{k=1}^{K} N_k = N~.
\ee

The EM algorithm proceeds as follows:

\begin{enumerate}
\item Choose the number of kernels, $K$.

\item Initialize the means ${\vec \mu_k}$, covariance matrices $\Sigma_{k}$ and mixing coefficients $\pi_k$ of the $K$ kernels.

\item Evaluate the responsibilities using the current set of parameter values (see Eq. \ref{resp}).

\item Re-estimate the parameters using the current responsibilities (use Eq. \ref{nk} for $N_k$):
\be
{\vec \mu_{k}}^{new} = \frac{1}{N_k} \sum_{n=1}^{N} \gamma(k \rightarrow n) {\vec x_n}~,
\ee
\be
\Sigma_{k}^{new} = \frac{1}{N_k} \sum_{n=1}^{N} \gamma(k \rightarrow n) ({\vec x_n}-{\vec \mu_{k}}^{new})({\vec x_n}-{\vec \mu_{k}}^{new})^{T}~,
\ee
\be
\pi_{k}^{new} = \frac{N_k}{N}~.
\ee

\item Evaluate the log-likelihood of this configuration:
\be
\ln L = \sum_{n=1}^{N} \ln \lbrace \sum_{k=1}^{K} \pi_k^{new} F({\vec x_n} \mid {\vec \mu_k}^{new}, 
\Sigma_k^{new}) \rbrace~.
\ee
and check for the convergence of the model parameters or of the likelihood. If convergence is not achieved, return to step $3$.
\end{enumerate}

Step $3$ is called the E (for expectation) step, while step $4$ is called the M (for maximization) step. In the case of non-Gaussian kernels, more sufficient statistical estimators can be computed during the 
maximization step. Whatever the kernel $F$, it can be shown that each E-M steps cycle increases the log-likelihood until it reaches a local or global maximum. One can hope to find the global maximum when starting an EM procedure with different initializations (step $2$). This is a reasonable approach if $K$ is small. The search for the global maximum is a much more difficult problem when $K$ is large, except if the likelihood landscape has only a small number of local maxima.

It will be stressed later that the finding of a strong local maximum of the likelihood function does not necessarily imply that the EM algorithm converged to the best solution. The next section will review some geological and seismological arguments that may help us constraining the best choice for the kernel $F$, which will, perhaps surprisingly, be found to be nothing but a multivariate Gaussian distribution (see section \ref{sschoice}).


\section{Geological and seismological constraints on the shape of kernel $F$ \label{ssclustfaultzone}}

We aim at extracting features labelled as `faults' from the spatial structure of an earthquake catalog. To perform this task, the EM algorithm first requires a plausible seismicity model of a fault, {\it i.e.}, a mathematical kernel that describes the spatial probability density function of events triggered by a single fault. The simplicity of the EM approach mainly relies on the existence of a generic kernel $F$ which, when properly rescaled, describes 
as accurately as possible what a fault is. From now on, our first assumption is that such a kernel exists. Yet, there is no reason for this kernel to be unique, as there may be different families of kernels able to describe different families of faults.

The vast majority of pattern recognition studies which use an EM approach consider very simple and highly symmetric kernels such as the multivariate Gaussian, without bothering to justify such a choice other than using
the argument of computational simplicity. In our opinion, the definition of this kernel should derive instead from the prior empirical or physical knowledge which is available concerning earthquake clustering on faults, as well as from their geometry or dynamics. We thus present below some rumination about the general properties that $F$ should possess for the specific application of data clustering to the determination of faults from earthquake catalogs. These considerations will take account of the numerous, and sometimes contradicting, empirical estimations of the statistical properties of earthquake and fault catalogs. The final determination of $F$ will be presented and justified in section \ref{sschoice}. It will take account of the fact that both knowledge and absence of knowledge have to be considered when defining $F$, thus requiring to couple the observations compiled in the present section with some information theoretical arguments.

\subsection{Present knowledge about the kernel $F$}

Notwithstanding the large amount of work and significant progress obtained on the structural and physical description of earthquakes and faults, there is no consensus on such a kernel since no theoretical model is successful in predicting the mechanical behavior of the brittle crust over the many different time scales necessary to take into account when going from earthquakes to faults. There are several noteworthy attempts to develop 
numerical simulations that model the progressive development and organization of fault networks
by the cumulative effect of earthquake activity (see e.g., Miltenberger et al. (1993), Sornette et al. (1994; 1995) and
Lyakhovsky et al. (2001)), but none of them has been seriously validated on natural data. 
We thus have to examine how empirical observations can help better constrain and define the kernel $F$.

The ideal situation would consist in observing a single and isolated fault and record its seismic activity over a long period of time. We could then provide a fit to the observed spatial distribution of events, look at its dependency on various parameters such as the applied stress field, estimate the influence of major lithological interfaces such as the free surface or the Conrad or Moho discontinuities on the shape of the resulting cluster, and so on.

Unfortunately, such direct observation is an illusion, as any tectonic domain features a large collection of active faults that mechanically interact, connect, or even cross each other. Fault networks (as observed by field geologists) can then feature very complex structures with variable strikes and dips, while those quantities may also vary spatially within the same fault zone. There is thus no hope to use the simple one-body approach.

In the next paragraphs of this section, we shall first focus on the representation of faults by seismologists and structural geologists, as well as on the probable influence of the time-scale of observations on these definitions. We shall then examine the short-term and long-term statistical properties of individual faults and ruptures. The case of fault networks and earthquake catalogs as a whole will also be briefly reviewed.

\subsection{Influence of the time scale of observations \label{faultdef}}

\begin{figure}
\noindent\includegraphics[width=15cm]{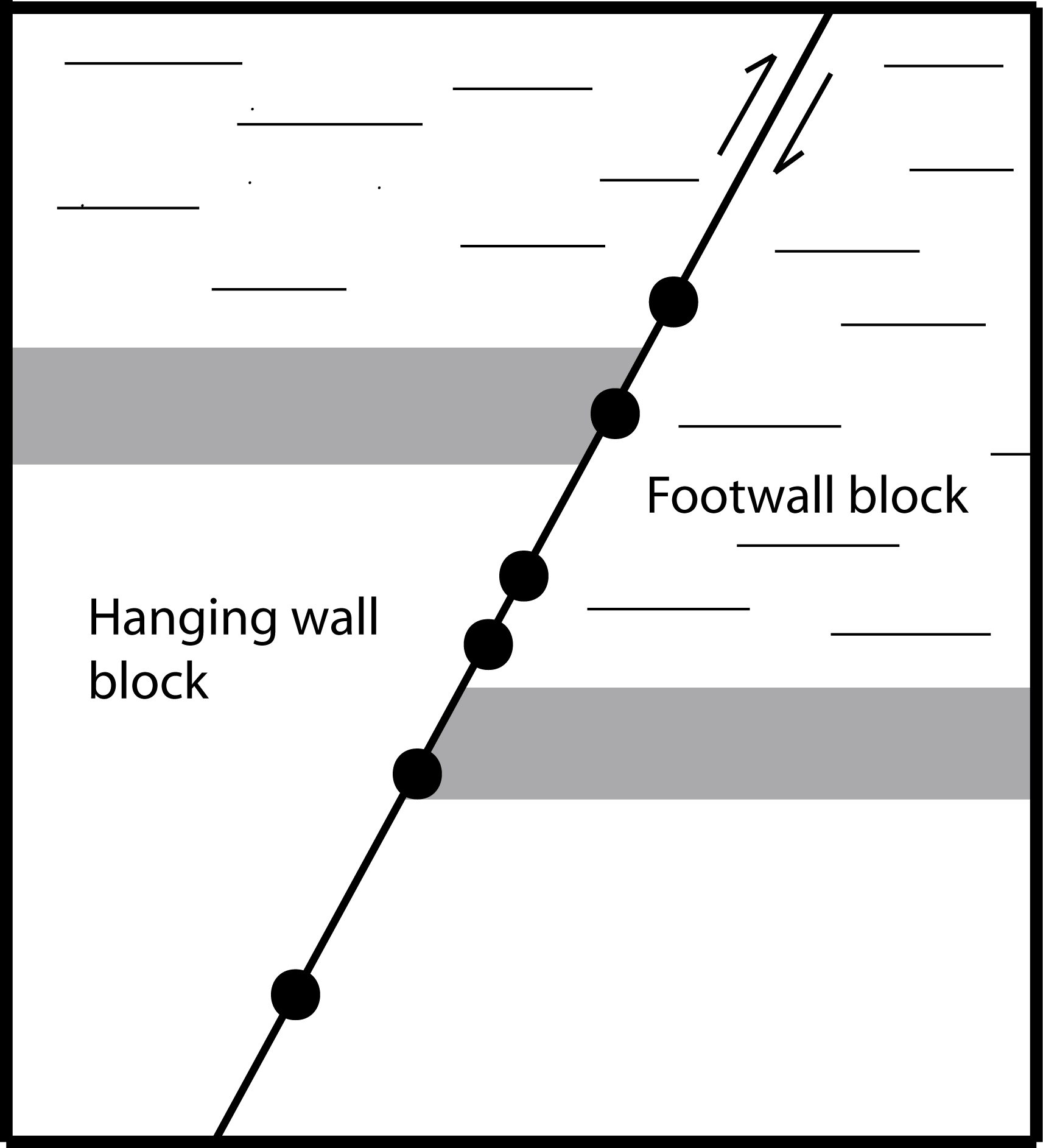}
\caption{\label{faultsismo} A fault as usually modeled by seismologists (dark oblique straight line), together with its potential seismicity (black circles).}
\end{figure}

The nature of the object defined as a fault may change with the time scale of observation. 
Consider first an earthquake as viewed by a seismologist. The typical time scale of a single detectable event (say, with magnitude $M \ge 1$) ranges from $\simeq 0.01$s to $\simeq 5$min, and corresponds to the duration of the rupture propagation. The geometry of the rupture process associated with such an event is generally considered as a perfect plane or a slightly curved surface, defined as the boundary between a footwall and a hanging wall (see Fig. \ref{faultsismo} for the case of an earthquake with thrust focal mechanism). Segmentation features along the earthquake rupture are sometimes mapped but their reliability is questionable, as they sometimes reveal more some interaction processes with the free surface, and may therefore be not representative of rupture at seismogenic depths. 
At the shortest time scales of an earthquake rupture, a fault thus seems to be a rather simple object whose identification and quantification is made rather straightforward by direct observation and/or sophisticated geophysical techniques. Fig. \ref{faultsismo} also shows the location of hypothetical events located on such a fault. They should thus define perfectly planar clusters, so that $F$ should look like a planar segment, corresponding to a flat kernel. Potential irregularities of the fault plane geometry as well as location errors would constrain $F$ to have a non-zero yet small thickness. If the fault was curved or clearly segmented, we could approximate it using several planar segments. According to a seismologist, a fault would thus be close to a completely deterministic euclidean structure which shape could be analytically determined. This view is in favor of a hard assignment methodology of clustering, yet one would have to take location errors into account. This can be done approximately as in Ouillon et al (2008), where it was assumed that the thickness of a cluster has to be smaller than the average earthquake location errors, or more elegantly, as in Wang et al. (2009), by using the expected square distance. The expected square distance is computed by using all distances from a plane to an earthquake location weighted by their full probability density function (pdf). In contrast, the euclidean distance used in Ouillon et al. (2008) takes account only of the location of the hypocenter (i.e. the most probable location).

\begin{figure}
\noindent\includegraphics[width=15cm]{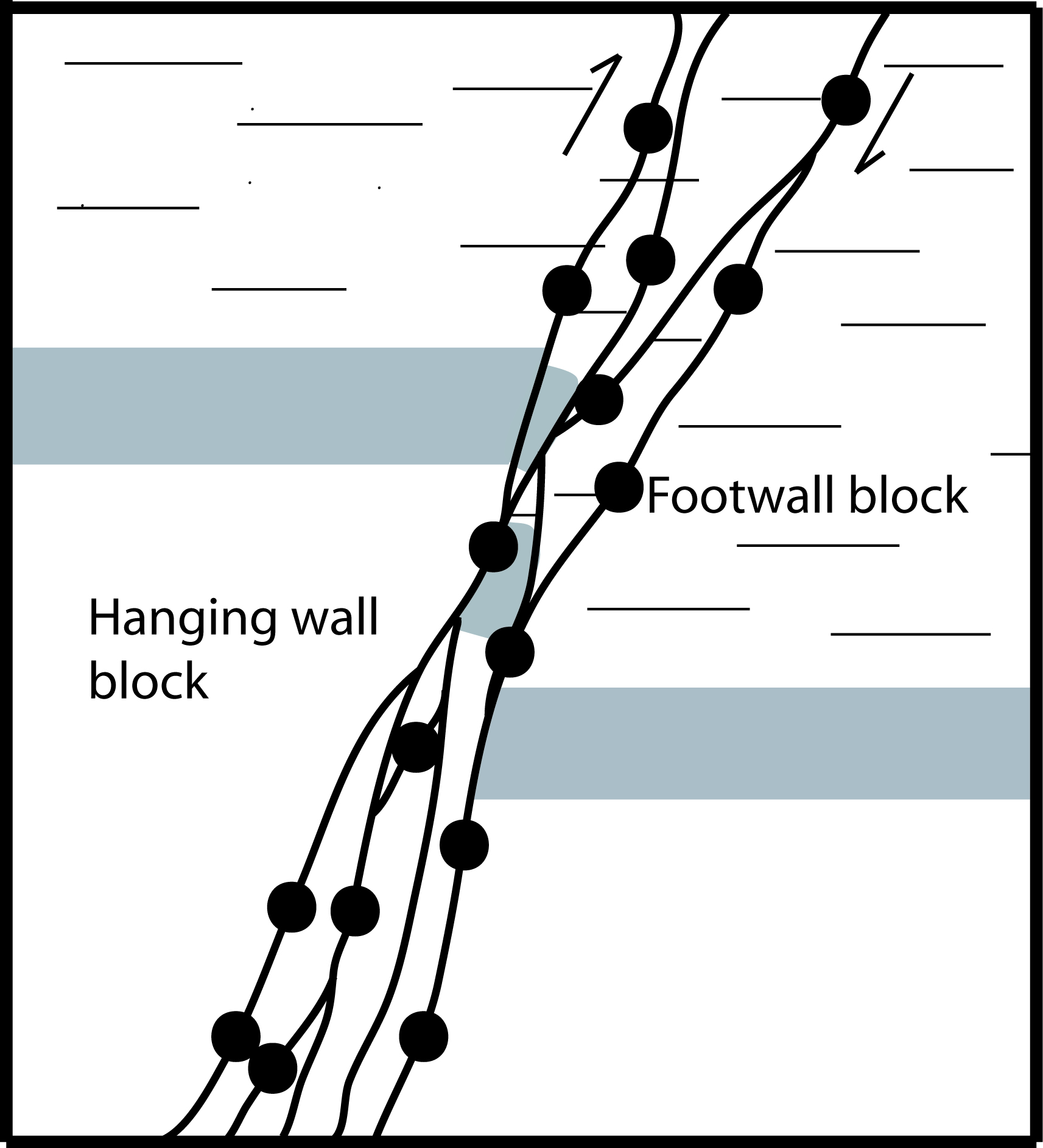}
\caption{\label{faultgeol} Schematic view of a fault as observed by a structural geologist (dark oblique lines), together with its potential seismicity (black circles).}
\end{figure}

Consider now a fault as viewed by a structural geologist. In this case, the relevant time scales typically range between $\simeq 10^5$~yrs and $\simeq 100$~Myrs. The rupture geometry appears as a more diffuse damage zone (with maximum intensity at its core), as very schematically depicted in Fig. \ref{faultgeol}. In this case, no mathematical description of its geometry is known, as many small-scale structures (often qualified as sub-faults) define a complex set of numerous interwoven and branching planes which are impossible to identify individually. The only solution is thus to consider this collection of sub-planes collectively, so that one has to deal with the fault zone as a whole. Such a complexity necessitates a stochastic description because, while the deterministic geometry varies along the fault, yet it may conserve the same statistical properties (such as fracture density). Fig. \ref{faultgeol} also shows a set of hypothetical hypocenters occurring on such a structure. This set does not define a plane but a rather fuzzy structure which has to be considered collectively as well. Some of the events may nucleate on different sub-faults, but they may partly share their complete rupture paths. The rupture geometry of a single event may be still simply approximated by a smooth surface, yet all possible rupture surfaces within a fault zone cannot be isolated individually and have to be collectively considered as defining the same entity. For the same reasons as above, this set of events has to be described using statistical tools, and this last example may help to better understand the need for using a kernel to describe the seismicity of such a tectonic structure. This suggests that, for intrinsic reasons, the kernel $F$ cannot be flat, and its thickness results from both the fuzzy structure of the fault zone and the earthquakes location uncertainties. Note that this model is compatible with the wide distribution of focal mechanisms of earthquakes thought to occur on the same fault (Kagan, 1992).

Recent observations within fault zones provide a slightly different picture (see Ben Zion and Sammis, 2003, for a review). It seems that major fault zones indeed appear as a diffuse damage zone (typically of a few hundreds of meters wide) within a more intact rock matrix, but the majority of slip across the fault seem to be accommodated within a very thin and finely crushed gouge zone called the principal fracture surface. Due to the mechanics of rock fragmentation within fault zones as well as indicated by field observations, this principal fracture surface develops rather early within the damage zone which then becomes mainly inactive, as most events nucleate and propagate along the principal fracture surface. This localization of slip is favored by strain weakening mechanisms, so that, on the long term, major faults can be considered as purely euclidean surfaces. This view is in agreement with works on strike-slip fault segmentation (see De Joussineau and Aydin, 2009) which show that the density of fault steps decreases as total displacement offset increases across the fault. No rigorous result is available about the structure of the active part of fault zones at depth, if 
indeed some activity occurs there. Robertson et al (1995) computed a fractal dimension in the range $D=1.8-2.0$ for relocated sets of hypocenters of aftershock sequences Southern California. Unfortunately, the observed self-similar scaling holds for scales beyond $1$~km, i.e. larger than the typical width of the damage zone. We shall thus assume that, in the long term view of faulting, earthquakes may occur within the damage zone surrounding the principal fracture surface.

As typical earthquake catalogs span only a few decades, it would seem that the seismologist's viewpoint should be the most relevant to constrain the shape of the kernel $F$. However, taking into account the empirical evidence and theoretical arguments that the growth of a fault is a self-similar or self-affine process in space as well as in time (Scholz et al., 1993; Scholz, 2002), we are led to recognize that the time scales involved in the statistical analyses of instrumental catalogs stand half-way between the seismologist's and the geologist's viewpoints. It follows that the kernel $F$ should thus feature characteristics that are common to both descriptions, and be as general as possible for all other properties. For example, $F$ should define an anisotropic kernel that allows one to identify clusters with different lengths, widths, strikes and dips. But their thicknesses (defined as the size of the cluster perpendicularly to its best fitting plane) should not be constrained to be zero so that we leave the possibility to detect faults as seismologists or structural geologists model them.

We face another problem along the time dimension, as faults grow with time (and earthquakes are thought to be part of this process), so that a given mathematical kernel may be valid to describe the seismicity triggered by a given fault only for a limited span of time. Thus, it may be possible that each fault would be described by a different kernel because they are at different stages of maturity. In order to take account of that possibility, the EM algorithm should be modified so that the mixture consists of the superimposition of kernels possibly belonging to different families (Gaussian, exponential, power-law, ...). This would increase drastically the dimension of the solution space, hence the complexity of the landscape in which to search for the maximum of the likelihood function. We thus discard that option in the present work and shall now review the basic quantitative geometrical properties of individual geological ruptures at different time scales.

\subsection{Direct observations of earthquake and damage distribution around a fault}

Statistical studies of natural seismicity on and around a fault, and of the structure of damage within and around fault zones,
are relatively scarse. Most of them focus on brittle deformation features perpendicular to the observed fault planes, i.e., in the direction that corresponds to the thickness of the clusters we aim at identifying.

\subsubsection{Distribution of damage across faults\label{sdamage}}

Vermilye and Scholz (1998) studied the density of micro-cracks as a function of the distance from the fault for two strike-slip faults of different sizes in the same locality and rock type. They concluded that the crack density at the fault core is independent of the size of the fault, and decays exponentially away from it. It seems that this claim comes from a misinterpretation of the plots (a confusion
of the two axes), as their plots show unambiguously that the decay is in fact logarithmic, not exponential. Extrapolating this dependence implies that the density of micro-fractures would drop to zero or to a background level at a finite distance from the fault, defined as the typical process zone width. This typical process zone width was found to be linearly related to the length of the fault (using additional data of different types). More recent data from the Caleta Coloso fault in Chile published by Faulkner (2006) provide a different result, according to which the density of micro-cracks is found to decay exponentially with distance from the fault core. Other data (compiled in Fig. 3.14 in Scholz (2002)) suggest that the gouge thickness within a fault scales with its length.

Those observations suggest that, if the seismicity rate depends on the local density of damage, the kernel $F$ should decay in the direction normal to the fault with tails that are thin enough so that all its statistical moments remain finite. This result allows us to choose a kernel which may be a function of the covariance matrix. The typical thickness of a cluster could be taken to depend linearly on its length, but the proportionality constant is not known.

\subsubsection{Distribution of earthquakes across faults}

Liu et al. (2003) showed that the spatial dispersion of aftershock epicenters normal to the direction of the Landers main rupture could not be explained by location errors alone, and that the event density decays exponentially with distance away from the fault (see their Fig. 6). In the same figure, we can also notice that the decay on both sides of the fault is approximately symmetric. These data must be considered with caution as they concern aftershocks that may occur on distinct faults, and not on the main fault zone. They also reveal the dynamics of the fault pattern on rather short time scales. In any case, this suggests that all statistical moments of the earthquake distribution in the direction normal to a fault are finite (see section \ref{sdamage}).

In a recent work, Hauksson (2010) performed a similar analysis of the seismicity occurring close (and in the direction normal) to major fault zones in Southern California. This study shows a varied behavior depending on how data are selected. The first case deals with the full set of events in the catalog that occur close to a given fault. Earthquake density shows a peak that sometimes coincides with the core of the fault zone, or is sometimes clearly offset on one of its side (note that the earthquake location relative to the fault takes account of the dip of the fault). The average shape of the decay of the earthquake density with distance from the fault displays a zone with a constant density level within $2$~km of the fault core, then decays as a power-law at larger distances with an exponent $\simeq 1.5$ to $2$. When selecting only events that are aftershocks of a larger event (such as Landers), the exponent is significantly larger ($\simeq 3$), thus suggesting a more localized kernel for short time scales. The density decays over distances larger than $10$~km, thus defining a wide {\it seismic damage zone}. Such a large scale suggests that events are likely to occur on different individual faults, a conclusion supported by the fit of the Landers aftershock sequence by Ouillon et al (2008), who showed that $16$ planes are necessary to explain the data at the scale defined by average location errors. The data of Hauksson (2010) thus cannot be used to constrain $F$ directly. His results suggests however that finite statistical moments of second order may be computed in the direction normal to the clusters. This feature will be used in section \ref{sschoice}.

\subsubsection{Distribution of earthquakes and damage along faults} 

The structure of fault zones is very complex along their strike (Martel et al, 1988; Martel, 1990; Willemse et al, 1997). This complexity mainly stems from the interactions among sub-faults which grew and finally connected to compose a larger fault. Fault zones thus display bends, pull aparts and step-overs at every scale, which appear as a segmentation process. This heterogeneity of the fault zone thus ensures that earthquakes do not occur uniformly over a simple plane. No model is available to describe this heterogeneity so that we cannot put any constraint on $F$ in that direction. Note that the disorder due to fault linking process certainly also holds in the vertical direction (Nemser and Cowan, 2009), hence contributing to the complexity of the distribution along the fault dip as well. Plotting histograms of the number of events as a function of depth for individual faults or for entire regions, Scholz (2002) showed that the organization of seismicity is quite complex and that the observed distribution sometimes displays multiple modes. Indeed, earthquake occurrence is governed by several mechanical factors that influence the frictional and rupture properties of rocks, that are all depth-dependent: lithostatic pressure, fluid pressure, as well as temperature (hence chemical processes such as stress corrosion) are thought to increase with depth. As these parameters also influence the bulk rheology of rocks, deformation mechanisms depend on depth too. The existence of horizontal rheological boundaries as well as of the free surface also favor a symmetry breaking of the seismicity patterns along the vertical direction. Yet, there is no clear model that is able to predict the shape of the earthquake distribution as a function of depth. Our choice for $F$ should thus not be too much constrained along the vertical direction. Note that the finite thickness of the seismogenic crust ensures a finite (if not well-defined) variance of the set of associated events along the dip of the fault. We have no proof that a well-defined and finite variance also holds along the strike of the fault.

\subsubsection{Implications for $F$}

The observation of damage around and along faults both on the long and short time scales shows that it is reasonable to assume that the distribution of seismic events within an individual cluster possesses finite second order moments, i.e. a variances, in any direction. It thus makes sense to consider a kernel $F$ which depends on the covariance matrix of the earthquakes' locations.

\subsection{Dynamics of a single fault}

A few other field observations may help us to better define the shape of $F$. They mainly deal with the observed cumulative displacement profile along a fault (which results from the successive occurrence of earthquakes on that fault since its formation), as well as with the coseismic displacement over a rupture plane during a single seismic event. These observations are of prime importance as displacement gradients control the stress field, hence the probable location of earthquakes. Scholz (2002) showed that cumulative displacement profiles along faults were self-similar, {\it i.e.} that they were almost identical when normalized by the fault length. Such a scaling law has also been documented for individual earthquakes, as coseismic displacement is observed to be proportional to the size of the rupture (see also Scholz, 2002). As the overall shape of the displacement field does not depend much on the scale of observation, this suggests that the kernel $F$ should be independent of the fault size. Thus, there is no need for different kernels to describe distinct faults. However, as the proportionality coefficient relating size and displacement differs between the single earthquake (where it is in the range $10^{-4}-10^{-3}$) and the finite fault case (where it is in the range $10^{-2}-10^{-1}$), the aspect ratios of $F$ should not be considered to be constant from one fault to another.

In a recent work, Manighetti et al. (2005) noted that cumulative (resp. incremental) displacement profiles along faults (resp. during an earthquake) display a strong spatial asymmetry that was not taken into account in the previously mentioned scaling laws. They also noted that this asymmetry is correlated with the presence of heterogeneities within the Earth's crust, and that displacement maxima occur close to those heterogeneities, then considered as potential nucleation locii of rupture. Ben Zion and Andrews (1998) observed such an asymmetry when modeling the propagation of a rupture along an interface between materials with different mechanical properties (which could be the damage zone on one side and the more intact host rock on the other side in the case of natural faults). This asymmetry of displacement profiles suggests that the seismic activity triggered by faults should also be asymmetric (yet we have no prior idea of how the former asymmetry should translate into the latter), so that the mode and barycenter of $F$ should not coincide. We shall check in section \ref{sschoice} whether it is possible to choose $F$ in a rigorous manner so that it displays this property.

\subsection{Geometrical self-similarity}

Many studies document the geometrical self-similarity of faults and earthquakes catalogs, which made concepts such as fractality or multifractality very popular in Earth Sciences and tectonics. Spatial self-similarity relies on two different kinds of statistics: the full size distribution of single objects (like fault or earthquake rupture length) and the scale dependence of the statistical moments of the spatial density distribution of objects (quantified by a set of generalized fractal dimensions or a multifractal spectrum).

The study of faults and of earthquake ruptures systematically shows that they are not single-sized objects, but that their lengths are distributed over a very broad interval. Different types of size distributions have been proposed, but it seems that a majority of studies agree with a power-law. This means that the rupture length probability density distribution scales as $P(L) \simeq L^{-a}$, with most observations suggesting that $a$ is universally close to $2$ for faults (Sornette and Davy, 1991) and to $3$ for earthquakes. This implies that the generic kernel $F$ should be scalable so that it can fit with clusters of different sizes (note that the EM algorithm could perfectly be defined without those degrees of freedom, considering a kernel $F$ that is independent on the covariance matrix, for example). 

The distribution of earthquakes and faults is not uniform, but features strong, intermittent bursts of spatial density (see for example Ouillon and Sornette, 1996). This complexity can be generally summarized in a well-defined multifractal spectrum $(\alpha , f(\alpha))$. This notation means that any box centered on an object (a fault or an earthquake) features a cumulated mass of similar objects that scales as a power of the box size, with exponent $\alpha$. This exponent, coined {\it singularity strength}, may vary in space so that set of points with the same value of $\alpha$ is itself distributed in space with a well-defined fractal dimension $f(\alpha)$. Indeed, a multifractal distribution can be seen as a complex mixture of self-similar subsets of data points with similar singularity strengths. However, this mixture is much more subtle and difficult to model than the one defined in section \ref{smixture}, especially because it is fundamentally nonlinear. The multifractal formalism implies that the local exponent $\alpha$ varies in space in a very complex manner, so that the shape of the kernel $F$ should also be able to display this kind of spatial variability. 

When dealing with faulting, multifractality quantifies the way small faults aggregate into clusters that behave as larger faults, themselves grouping into larger clusters defining some other faults, and so on (Ouillon et al., 1995; 1996). Such clusters exist at all scales, which makes more complex the identification of individual faults. The same holds for the clustering of earthquakes.
Many studies suggested that, within a fault network, the average $\alpha$ value (corresponding to the so-called correlation dimension) should be close to $1.7$ on $2D$ maps of fault traces. Ouillon et al. (1995; 1996), considering the network of fault traces observed in Saudi Arabia, and taking account of several biases that can alter the multifractal spectrum (Ouillon and Sornette, 1996), suggested an average value closer to $1.8$, so that a fractal dimension of $2.8$ may hold in $3D$, if one assumes isotropy (which is certainly a strong assumption). In the case of earthquakes, the correlation dimension seems to be close to $2.2$ (Kagan and Knopoff, 1980).

Despite the fact that the exponents measured on fault and earthquake catalogs may vary from region to region, it seems that their difference is significant, so that none of them can be reliably used to constrain the shape of the kernel $F$. Anyway, the above scaling laws would suggest that $F$ should exhibit power-law tails, with exponent values possibly informed by the multifractal spectrum properties mentioned above. 
The problem with this approach is that multifractality refers to a global property, in general isotropic, to which all events contribute even if they belong to different faults. The knowledge of the exponents quantifying the singularity strengths of the multifractal spectrum helps only if the underlying distribution is invariant under rotation, but tectonic deformations clearly display privileged directions. Thus, we should use a kernel behaving as a power-law with an exponent that possibly varies with azimuth. Unfortunately, no data is yet available that could constrain the definition of such a kernel. 

In addition, power law kernels have the inconvenience of requiring large scale or small scale cutoffs (depending on the local value of $\alpha$), to remove the pathological behavior associated with the mathematical singularity of the power law which impacts on the calculation of statistical moments such as the mean or the variance. These quantities are necessary for estimating the position and size of a cluster, and ultimately control their computed values. The values of those cutoffs, and their possible origin, are still largely unknown (Sornette and Werner, 2005a).

Maybe the most important feature of the multifractal formalism is that it also implies that, whatever the scale, each event is the center of a self-similar cluster, so that, strictly speaking, one should set $K=N$ in the EM algorithm. In a sense, this is the approach that is used when fitting earthquake sequences using cascade models of seismicity such as the ETAS model.  In cascade models of seismicity, each event is the center of a component of the cascade in time as well as in space. Unfolding the cascade can be performed by using model-dependent approaches (such as Zhuang et al., (2002; 2004) who use the ETAS model) or model-independent approaches (such as Marsan and Lenglin\'e, 2008). The former assumes an analytical dependence of part of the kernel parameters on the magnitude of the source event, so that it deals with a marked process. The latter does not assume any analytical form of the kernel, yet it has to be the same for all events, or to depend on their magnitude. In addition, these models assume linear superposition of the kernel contribution of each event to the triggering intensity. In the end, both methods separate all events in two broad families: a set of independent (or background) events, and a set of triggered events (``aftershocks'' in the standard language; see Helmstetter et al. (2003) for a detailed discussion of how to understand the concept of aftershocks in this context). Empirical evidence shows that the share of independent events decreases as the magnitude threshold of the considered catalog decreases (Sornette and Werner, 2005b). Thus, the set of independent events that are determined from such studies cannot be seriously considered as the seeds around which clusters are formed, i.e. as the potential centers of clusters. Indeed, as all events have to be considered as the center of a cluster, it does not help at all for a data clustering approach. ETAS-like models and their derived techniques search for probabilistic relationships of causality between events, which are not adapted for the goal of defining faults.

These arguments thus suggest that it is not obvious how to use the quantitative knowledge concerning the self-similarity of earthquake distributions in order to constrain the kernel $F$ that is needed to develop an efficient clustering algorithm. The problem is that the multifractal description does not naturally fit with the concept of data clustering with the goal of identifying intermediate scale structures (faults), essentially because the multifractal formalism is based on only two relevant characteristic scales: the whole catalog and the individual event.


\section{Gaussian mixture approach \label{sgaussmix}}

\subsection{Choice of the kernel \label{sschoice}}

In section \ref{ssclustfaultzone}, we have examined in details how the theoretical and observed geometry of earthquake clusters that emerge over various time and space scales could inform us on the shape of the kernel $F$ needed for the cluster analysis. Unfortunately, this synthesis of the available
knowledge suggests that a constructive approach to derive physically a plausible kernel may not yet be the most efficient one. In the present section, we show that our lack of knowledge can actually be used to deduce 
unambiguously a reasonable analytical form of the kernel $F$ kernel. 
The key idea is to quantify the impact of the lack of knowledge combined with what we know 
using the framework of information theory. As more and better data are incorporated in improved
statistical and physical models,
the proposed approach of combining these models  with information theory provides a general
framework to improve on the kernel that we propose now. Therefore, the choice
of the kernel proposed here should be considered only as a first approximation step, which is
open to significant improvement in the future.

In order to identify earthquake clusters, we need to organize their description. Given a spatial distribution of events assumed to belong to the same cluster, 
it is natural to characterize it by the hierarchy of its moments. The first one is the first-order moment, which defines the mean position noted $\vec \mu$ as it is a $3D$ vector. The mean position is a simpler quantity to compute than the mode of the distribution of event positions, especially if the distribution is multimodal within a cluster, a possibility that we cannot reject at present. The second quantity is the second-order centered moment of the distribution of event positions, otherwise known at the covariance matrix $\Sigma$ of the locations of the events. The tensor $\Sigma$ describes both the size and orientation of the cluster under study. 

Higher order moments of the distribution constrain further the distribution of events within a cluster. For example, in $1D$, the third and fourth order moments are related to the skewness and kurtosis of the distribution, which provide characteristic measures of the distribution asymmetry and distance to the Gaussian law. In the $3D$ case, the corresponding quantities are the $3rd$ and $4th$ order moments tensors, which are  much more delicate to estimate reliably from empirical data.  For the sake of simplicity and robustness, as measures of a given cluster, we will only use (i) its mean position $\vec \mu$ and (ii) its covariance matrix $\Sigma$. This implies that the distribution of events within a cluster falls off sufficiently rapidly (faster than a power law with exponent $3$) at large distances, such that the components of the covariance matrix are finite and well defined. This hypothesis is compatible with many of the various empirical laws of earthquakes and damage spatial density around faults mentioned in the previous section.

In Information theory, the lack of knowledge about a stochastic process is quantified by the entropy of the underlying distribution. In the discrete case, if event $i$ has a probability $P_i$ to occur (or if its relative frequency weight is $P_i$), the entropy of the distribution $P$ is defined as:

\be
H[P] = -\sum_{i} P_i ln P_i~.
\ee

A very good knowledge about the process means that $P$ is sharply peaked around a single value, and $H[P]$ will be close to $0$. In contrast, zero knowledge about the underlying process (just like playing dice with equivalent probabilities for all possible outcomes) corresponds to the situation in which all events $i$ have the same probability to occur. The entropy $H[P]$ then reaches its maximum value, which is $\ln N$ if the $N$ different events $i$ are equally possible. This formalism can be extended to the continuous case with the definition of the entropy of a spatial stochastic process with density $P$ as:

\be
H[P] = -\int_{\rm space} P({\vec x}) \ln P({\vec x}) d{\vec x}~.
\ee

Our problem is to determine the kernel $P$ that best models the shape of a cluster about which we know only (i) its mean position $\vec \mu$ and (ii) its covariance matrix $\Sigma$ (i.e., its size). This amounts to find the function $P$ which makes the entropy $H[P]$ maximum, in the presence of the constraints
on the value of $\vec \mu$ and $\Sigma$. Indeed, maximizing $H[P]$ means that we look for the kernel $P$ which assumes the minimum information beyond what we can determine from $\vec \mu$ and $\Sigma$. Maximizing $H[P]$ with fixed $\vec \mu$ and $\Sigma$ is solved in a standard way using the method of Lagrange multipliers. The unique solution for $P$ is the multivariate Gaussian distribution with mean $\vec \mu$ and covariance matrix $\Sigma$ (see Bishop, 2007).

This result is fundamental for the remaining of this paper and should thus be understood very clearly. We do not assume that all earthquake clusters are organized according to a multivariate Gaussian distribution. We just use the
fact that the Gaussian distribution is the most general {\it prior} distribution (which assumes the least additional information) to describe a cluster, provided only its location, size and orientation are determined.

For graphical purpose, we shall represent the determined multivariate Gaussian kernels using plane segments. To each Gaussian kernel mean $\vec \mu$ and covariance matrix $\Sigma$, we will show a plane centered at the position of the mean $\vec \mu$ of the Gaussian, while its orientation will be given by the principal axes of its covariance matrix $\Sigma$ corresponding to the two largest eigenvalues. Those two largest eigenvalues will also determine the size of the plane, following the convention of Ouillon et al (2008) (see section \ref{appsynthex}).

Section \ref{appsynthex} illustrates the Gaussian mixture approach on a synthetic example. It also elaborates a strategy to choose the number of Gaussian kernels used in the cluster decomposition, as it is one of the latent variables of the problem.

\subsection{Possible caveats \label{caveats}}

Having chosen the multivariate Gaussian kernel for $F$, the dataset studied below will be processed using the algorithm given in section \ref{ssem}. Recall that it uses a maximization of the likelihood as a criterion for the convergence towards a solution. 

There are some cases for which the validity of this approach can be questioned. Imagine a dataset featuring $N$ data points that we try to fit using $K$ Gaussian kernels. Also imagine that, within this dataset, we can define a small cluster featuring three points that are spatially isolated from the others. Assuming that $K$ is large, and depending on the initialization of the kernels, one may reach a situation where one of the kernels will account only for these three points and will thus literally collapse on them: this means that the covariance matrix of that kernel will depend only on the spatial locations of those points, all other data points having a vanishing influence on it. It thus follows that this covariance matrix will be singular, as the smallest eigenvalue will be zero, so that the likelihood contribution of this kernel will be infinite. The total log-likelihood will thus be infinite too, whatever the solution provided for the rest of the dataset. In that case, maximizing the likelihood does not have any meaning. More generally, this problem occurs in a $D$-dimensional space when the $D$-volume of a cluster is null. It is thus not appropriate for the identification of perfectly planar clusters in $3D$, a situation that is however unlikely in earthquake catalogs. Note that in the $3D$ case, a kernel can also collapse on a set of two events, or on a single datapoint.

Two solutions can be proposed to solve this problem. The first one consists in checking {\it a posteriori} that none of the kernels correspond to a singular covariance matrix. This can be easily done during the computations, and the corresponding kernel can then be removed from the set of kernels. The other (and best) solution consists in eliminating this possibility directly inside the dataset. In that case, we eliminate events that define local geometrical structures that are likely to be characteristics
of random catalogs instead of genuine geological features. This procedure will be described in section \ref{appmntlew}. 


\section{Application of the EM method with Gaussian kernels to a synthetic example \label{appsynthex}}

\subsection{Generating the dataset}

\begin{figure}
\noindent\includegraphics[width=15cm]{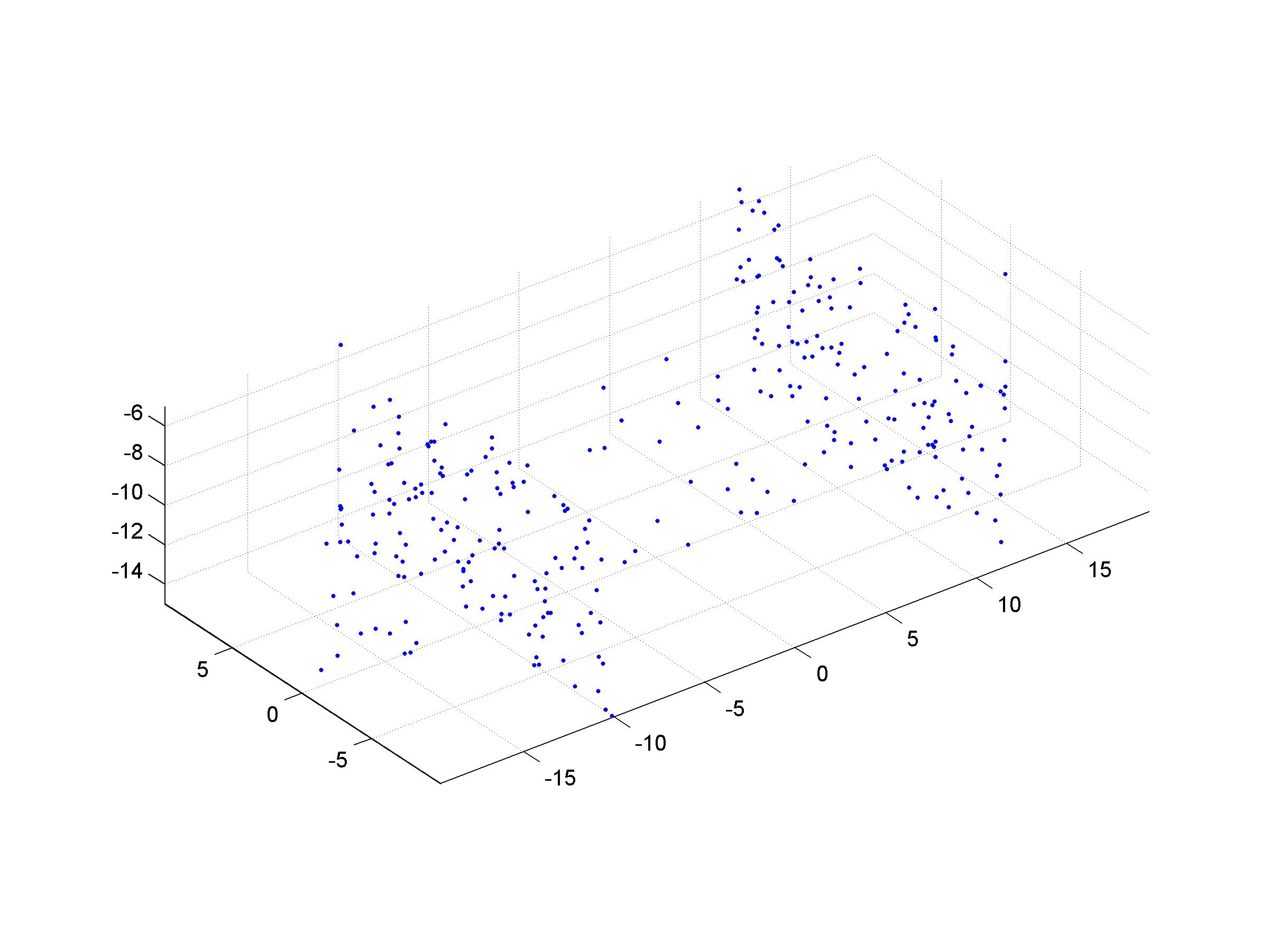}
\caption{\label{synth_0_plane} The synthetic example analyzed in section \ref{appsynthex}. See Table $1$ for its imposed geometrical parameters.}
\end{figure}

The synthetic example we present here is similar to the one we considered in Ouillon et al. (2008) for the sake of comparison. Events are located over a set of three vertical faults. Here, $\lambda$ stands for longitude, $\phi$ stands for latitude, and $z$ stands for depth. The center of the first fault is located at $\lambda=-0.01^{\circ}$ and $\phi=0$. The center of the second fault is located at $\lambda=+0.01^{\circ}$ and $\phi=0$. Both faults have their centers at $z=10$~km, while they are strictly parallel with a zero strike, vertical dip, length of $20$~km and width of $10$~km. The third fault is normal to the former ones, and its center is located at $\lambda=0$, $\phi=0$, $z=10$~km. Its width is $10$~km while its length is $40$~km. We locate randomly (using a uniform distribution) some events over each plane. The first two faults carry $100$ events each, whereas the third one carries $200$ events as it is twice longer. There are thus $400$ events in total. A random deviation to each event location is added in the normal direction to the plane it belongs to, mimicking location errors. This deviation is chosen uniformly within $[-0.01 {\rm km} ; +0.01 {\rm km}]$, so that it is comparable with relative location uncertainties in real relocated catalogs. This dataset is shown in Fig. \ref{synth_0_plane}. We assume that the user measures the position of each event, which is determined within an uncertainty of $\pm 0.01$~km in each spatial direction. The number of faults and their geometrical properties are not known and the goal of the exercise is to demonstrate how good is the EM method with Gaussian kernels to recover these faults and their properties.

\subsection{Fitting the data}

Our algorithm uses the following variables and notations. The kernels will be noted $G$ for Gaussian. A given kernel $i$ of a set of $K$ kernels is denoted $G_i^K(\vec{x} \mid {\vec\mu_i^K}, \Sigma_i^K)$. The index $i$ of the kernels runs from $1$ to $K$. For each kernel, the corresponding covariance matrix elements will be noted $\sigma_{i,mn}^K$, where indices $m$ and $n$ stand for $x$, $y$ or $z$. Each covariance matrix can be diagonalized, and its eigenvalues are noted $\sigma_{i,j}^K$, where $j=1$ for the largest eigenvalue, $j=2$ for the intermediate eigenvalue and $j=3$ for the smallest eigenvalue. The corresponding eigenvectors are noted ${\vec v_{i,j}^K}$. Following Ouillon et al (2008), we assume that the size of the fault in any of its eigendirections is $\sqrt{12}$ times the standard deviation of its associated seismicity in that direction.

In a first step, we fit this dataset using a single Gaussian distribution, i.e., we start with $K=1$. The dataset is first divided into two distinct subsets controlled by a probability $p$ such that $0 < p < 1$. Each event in the whole catalog has a probability $1-p$ to belong to a first set called the training set, while it has a probability $p$ to belong to the validation set. For this example, we set $p=0.1$ which means that, on average, about 40 events will belong to the validation set while about 360 events will belong to the training set. For each event, we thus draw a random number between 0 and 1, compare it to $p$ and decide whether the datapoint belongs to the validation set or to the training set. After selecting those two subsets, we change the spatial location of each event by adding a random perturbation similar to the one which has been used when generating the catalog (in order to simulate location uncertainties). Both subsets can then  be viewed as independent realizations of the dynamics of the same underlying fault pattern. We then fit the training set using only one cluster, whose solution allows to compute the likelihood of the obtained configuration. The log-likelihood is then normalized by the number of datapoints in the training set, so that we obtain a mean likelihood per datapoint (which we herafter refer to as $L_{pd}$). Using the same optimal kernel parameters we obtained for the training set, we now compute the $L_{pd}$ of the validation set conditioned to that kernel. In general, this likelihood will be smaller than the former. We are thus left with two $L_{pd}$'s conditioned to the same kernel: one for the training set, the other for the validation set. We perform this procedure ten times, selecting different training and validation sets for a fixed value of $p$, perturbing randomly their location according to spatial uncertainties, so that we can provide an average value and error bars for the $L_{pd}$'s in the case $K=1$.

We then add a second Gaussian kernel, which amounts to fit the data with two kernels $G_1^2(\vec{x} \mid {\vec \mu_1^2}, \Sigma_1^2)$ and $G_2^2(\vec{x} \mid {\vec \mu_2^2}, \Sigma_2^2)$. We call this operation {\it kernel splitting} as the primary kernel is split into two secondary kernels (the primary kernel is simply the one obtained in the last of the ten fits for $K=1$). The initial positions and covariance matrices of the secondary kernels are not chosen at random but depend on the properties of the primary kernel. We choose them such that:

\be
{\vec \mu_1^2} = {\vec \mu_1^1} + \frac{\sqrt{3}}{2}\sqrt{\sigma_{1,1}^1} {\vec v_1^1}
\ee
\be
{\vec \mu_2^2} = {\vec \mu_1^1} - \frac{\sqrt{3}}{2}\sqrt{\sigma_{1,1}^1} {\vec v_1^1}
\ee
\be
\sigma_{1,1}^2 = \sigma_{2,1}^2 = \frac{1}{4}\sigma_{1,1}^1
\ee
\be
\sigma_{1,2}^2 = \sigma_{2,2}^2 = \sigma_{1,2}^1
\ee
\be
\sigma_{1,3}^2 = \sigma_{2,3}^2 = \sigma_{1,3}^1
\ee

\begin{figure}
\noindent\includegraphics[width=15cm]{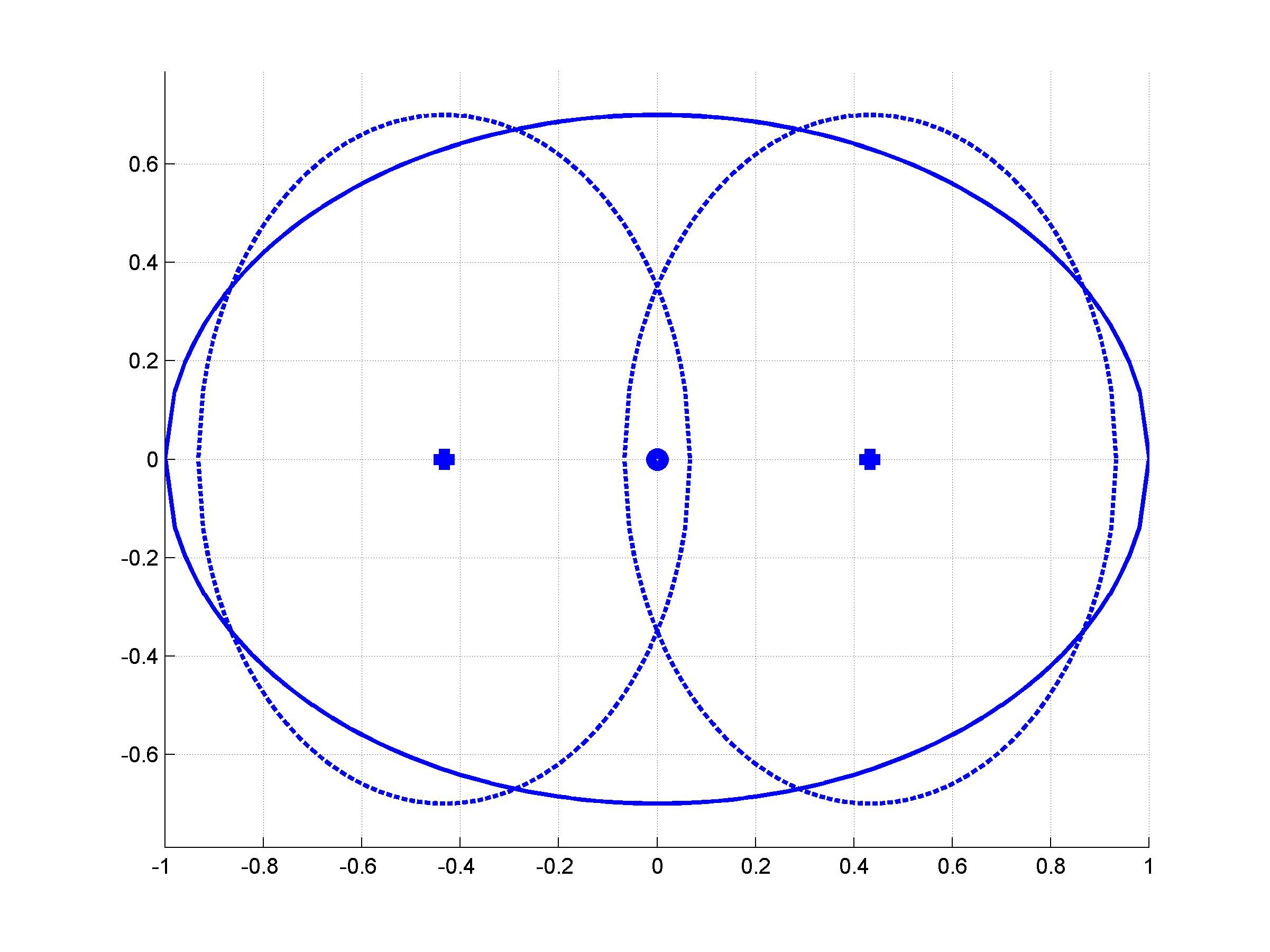}
\caption{\label{splitgauss} A $2D$ multivariate Gaussian kernel, represented by its 
one-standard deviation uncertainty ellipse using a continuous line and centered on the full circle at $(0,0)$, is replaced by two smaller-scale kernels represented by dashed lines and centered on thick crosses. We call this operation {\it kernel splitting}. See the main text for detailed explanations.}
\end{figure}

Figure \ref{splitgauss} shows the geometrical meaning of this substitution in the $2D$ case. In this figure, each multivariate Gaussian kernel is represented using its one-standard deviation ellipse (which becomes an ellipsoid in $3D$). The primary Gaussian kernel is drawn using a continuous line, and its center is represented with a full circle located at $(0;0)$. In this example, the standard deviation of the primary Gaussian kernel along the long axis is set to $1$, while its standard deviation along the short axis is set to $0.7$. This primary Gaussian kernel is then replaced by the two secondary kernels drawn using dashed lines, and centered on the full squares. For each of these secondary kernels, the standard deviation along the vertical axis of the figure is the same as the one of the primary kernel in that direction, i.e. $0.7$. In the horizontal direction of the figure, the standard deviation of each secondary kernel is set to half the standard deviation of the primary kernel in that direction, i.e. $0.5$. Both secondary kernels are respectively centered at $(\pm \sqrt{3}/2;0)$. This ensures that the global covariance matrix of the set of secondary kernels is identical to that of the primary kernel. We are thus left with a new configuration which has the same global covariance matrix as the primary one, except that it now features more degrees of freedom to fit the data, and the two secondary Gaussians can now be used as the starting configuration for the EM algorithm (while the primary kernel is removed from the set of kernels). In the $3D$ case, our procedure is also to split the primary cluster into two secondary kernels along its largest axis, and to leave untouched the other eigen-directions.

We then consider the full synthetic dataset, reassign each event either to the training set or to the validation set using the same probability $p$, and perturb their original spatial location with new random increments. Those two subsets are thus continuously changing while the algorithm proceeds, which allows a better exploration of the influence of all events, as well as of location uncertainties, in the fitting process. The training dataset is then fitted using $K=2$ kernels, and the $L_{pd}$ is computed for the best solution. The $L_{pd}$ of the validation set conditioned to the same set of kernels is also computed and stored. Once again, we generate ten different training/validation datasets to provide averages and error bars.

\begin{figure}
\noindent\includegraphics[width=15cm]{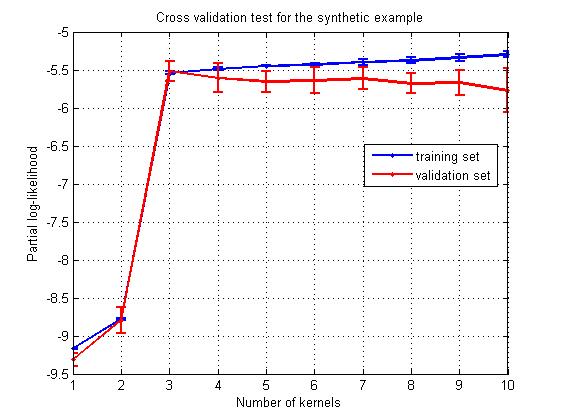}
\caption{\label{synth_crossvalid} Average values of the $L_{pd}$ and one standard deviation error bars for the training sets (upper curve) and the validation sets as a function of the number of kernels when fitting the dataset shown on Figure \ref{synth_0_plane}.}
\end{figure}

We then split the thickest kernel and start to fit the data with $K=3$ kernels, and so on. The thickest kernel is defined as the one which features the largest variance across its principal plane. For each value of $K$, we select a training dataset and a validation dataset, perturb the events spatial location, and compute the $L_{pd}$ for both datasets using the same kernels configuration. Fig. \ref{synth_crossvalid} shows a plot of the $L_{pd}$'s for the training and validation datasets versus the number of kernels used in the fit. One can check that the likelihood increases without any bound in the case of the training dataset: the more kernels we use to fit the data, the larger the likelihood. However, the increase is much slower for $K > 3$, indicating that increasing $K$ further may not be so efficient. The result is somewhat different for the validation set, which is an independent dataset. The $L_{pd}$ of the validation set first increases with $K$ in a manner similar to the training dataset: both datasets are similarly explained by the corresponding set of kernels. For values of $K$ larger than $3$, one can check that the $L_{pd}$ of the validation set decreases slowly, which means that adding more clusters to fit the training set provides less and less consistency with an independent dataset sampling the same underlying faults. This comes from the fact that increasing $K$ allows to fit smaller scale details of the training set whose geometry is more and more controlled by location uncertainties. This leads us to conclude that one should not use more than $K=3$ clusters to fit the dataset under study, which recovers
the true number of faults in this synthetic example.

\begin{figure}
\noindent\includegraphics[width=15cm]{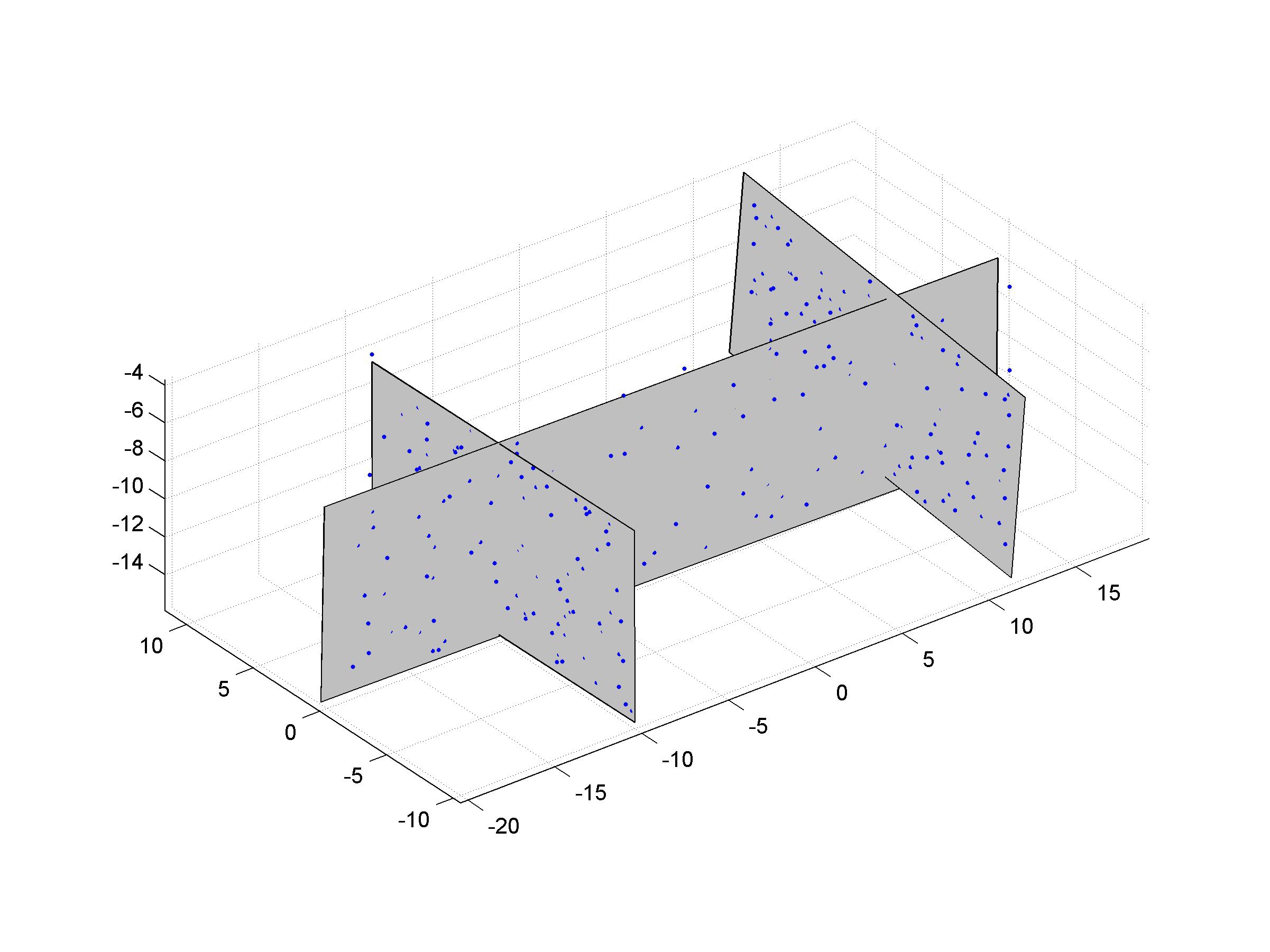}
\caption{\label{synth_3_plane} Data clustering of the synthetic set of events shown in Fig. \ref{synth_0_plane}, using three Gaussian kernels. See Table $1$  for their properties.}
\end{figure}

Fitting the full dataset with $K=3$ kernels, the EM algorithm converges to the pattern shown in Fig. \ref{synth_3_plane}. We check that it fits almost perfectly the planes used to simulate the dataset. Table $1$ shows the parameters of the synthetic simulation as well as the cluster parameters retrieved by this modified EM algorithm.

Of course, this methodology features a tuning parameter $p$ which value is chosen arbitrary (but a value $p=0.1$ is often assumed in cross-validation schemes). $p$ must not be too large, as the training set must feature as many events as possible to provide a reliable fit of the fault pattern itself (in order not to waste too many data). In the other hand, $p$ must not be too small, as the validation set may then feature too few events which would provide an unreliable value for the $L_{pd}$ of the validation set (we checked that if $p$ is too small, then this $L_{pd}$ may feature very strong fluctuations with $K$). The value $p=0.1$ is a
good compromise with a smooth behavior of the $L_{pd}$s as a function of $K$, together with a nice convergence of the algorithm when fitting the training dataset. We will thus use this value $p=0.1$ in our implementation below.
As the validation dataset is smaller than the training dataset, error bars are larger for the validation dataset. Using $p=0.05$ didn't change the results (even for the natural catalog analyzed in the next section), whereas $p=0.20$ increased the number of cases where kernels collapsed on three or less datapoints.


\section{Application of EM to the Mount Lewis aftershock sequence \label{appmntlew}}

\subsection{The Mount Lewis aftershock sequence}

The $M_L=5.7$ Mount Lewis earthquake occurred on March 31, 1986 on a right-lateral fault northeast of and oblique to the Calaveras fault in California (Zhou et al., 1993). Inversion of its centroid moment tensor suggests a right-lateral slip motion over a rupture plane striking $N353^{\circ}$ with a $79^{\circ}$ dip, and a $9$~km deep hypocenter. Zhou et al. (1993) studied the aftershock sequence in details, and deduced a main rupture size of typical radius of about $1.5 - 2.5$~km. The seismicity pattern shown in Figure \ref{mntlewisseq} is mainly due to the aftershocks following this event. Their focal mechanisms display strike-slip motion on steep faults, without obvious spatial relationships between them.

The seismicity in that area has been relocated by Kilb and Rubin (2002) who used waveform cross-correlation techniques to determine the relative positions of $2747$ events. While the relocated events were found to delineate a N-S trend at large scale, a more complex picture emerges at smaller scales. In particular, fine structures could be identified as forming series of E-W near-vertical faults of small dimensions ($\simeq 0.5-1$~km long), with separations between them as small as $\simeq 200$ m. Kilb and Rubin (2002)  proposed that these structures result from the growth of a relatively young, right lateral NS fault, whose displacement is accounted for by slips on secondary left-lateral EW faults. Near the location of the main shock, the structure is much simpler and close to a single NS plane. The occurrence of EW secondary faults is more obvious to the North of the main shock, and the lengths of those faults seem to increase with distance to the main shock. The secondary faults located South of the main shock seem to have more variability in depth and strike, so that the picture is somewhat more fuzzy. 

Kilb and Rubin (2002) argued that those EW faults might reveal a preexisting faulting anisotropy in the area. They also noticed that two EW clusters located respectively to the North and South of the main event were activated a few weeks later, suggesting a delayed growth process of the fault which generated the main shock. 

The catalog we used is the one of Waldhauser and Schaff (2008) which extends from January 1984 to May 2003. Yet, we only used events occurring after the Mount Lewis event, up to the end of the catalog. Note that this catalog features only events recorded by 6 or more stations.
Fig. \ref{mntlewisseq} (left column) shows that, if most of events define clusters that could be easily identified with the naked eye, some of them define much sparser structures (if any structure at all), that preclude the identification of any reliable underlying fault. Before applying our modified EM algorithm in order to identify automatically all significant clusters, we thus need to address the problem of uncorrelated seismicity.

This catalog features the $3D$ (relative) locations of events, as well as informations about their uncertainties. The epicentral uncertainty is given by the size (in km) of the axes of the $95\%$ error ellipse, together with the azimuth of its large axis. The depth uncertainty is given by the $95\%$ error interval (in km), so that we do not know the full covariance matrix of each event's location. In order to make computations more efficient, we determine the average of the three spatial standard deviations and assume that location uncertainty is isotropic. This simplified uncertainty sphere (of radius $\simeq 10$~m for most events) is then used to build the reshuffled training and validation sets, following the procedure described in section \ref{appsynthex}, assuming a Gaussian distribution of earthquakes hypocentral location.

\subsection{Pre-treatment to sort out clustered and unclustered seismicity \label{clustunclust}}

The synthetic example treated in section \ref{appsynthex} is idealized in several respects, one of them being that the events are all assumed to be associated to some fault, possibly different from event to event, and that all existing faults are sampled by a sufficiently large number of events. In real data sets, which necessarily include only a limited history of seismic recording, several events can be isolated or, more generally, not associated with any cluster. This does not mean that they did not occur on a fault but that these particular faults associated with these isolated earthquakes carry too few events (perhaps just one) to reveal their existence via any clustering reconstruction algorithm. One can say that these faults are under-sampled by the available seismic catalog.

Given the enormous complexity of faulting at all scales, the existence of these isolated events 
constitutes a problem in any clustering procedure, in order to keep small the number of reconstructed fault-clusters. Not taking into account the existence of this diffuse seismicity would lead in any clustering procedure to a proliferation of spurious faults. Indeed, for a set of isolated events, the clustering algorithm tends to partition that dataset into subsets of three events each, as this is the optimal solution for partitioning a more or less isotropic distribution with anisotropic clusters. Thus, in most cases, we would be confronted with the convergence problem raised in section \ref{caveats}, as some kernels will collapse on sets of three events or less, and the likelihood would diverge spuriously. The whole exercise would become meaningless and no insight on the relevant underlying fault network would be derived. We thus have first to identify the clustered part of seismicity, remove the uncorrelated seismicity and apply our EM procedure only to the clustered subset in order to identify and characterize individual clusters.          

\begin{figure}
\noindent\includegraphics[width=16cm]{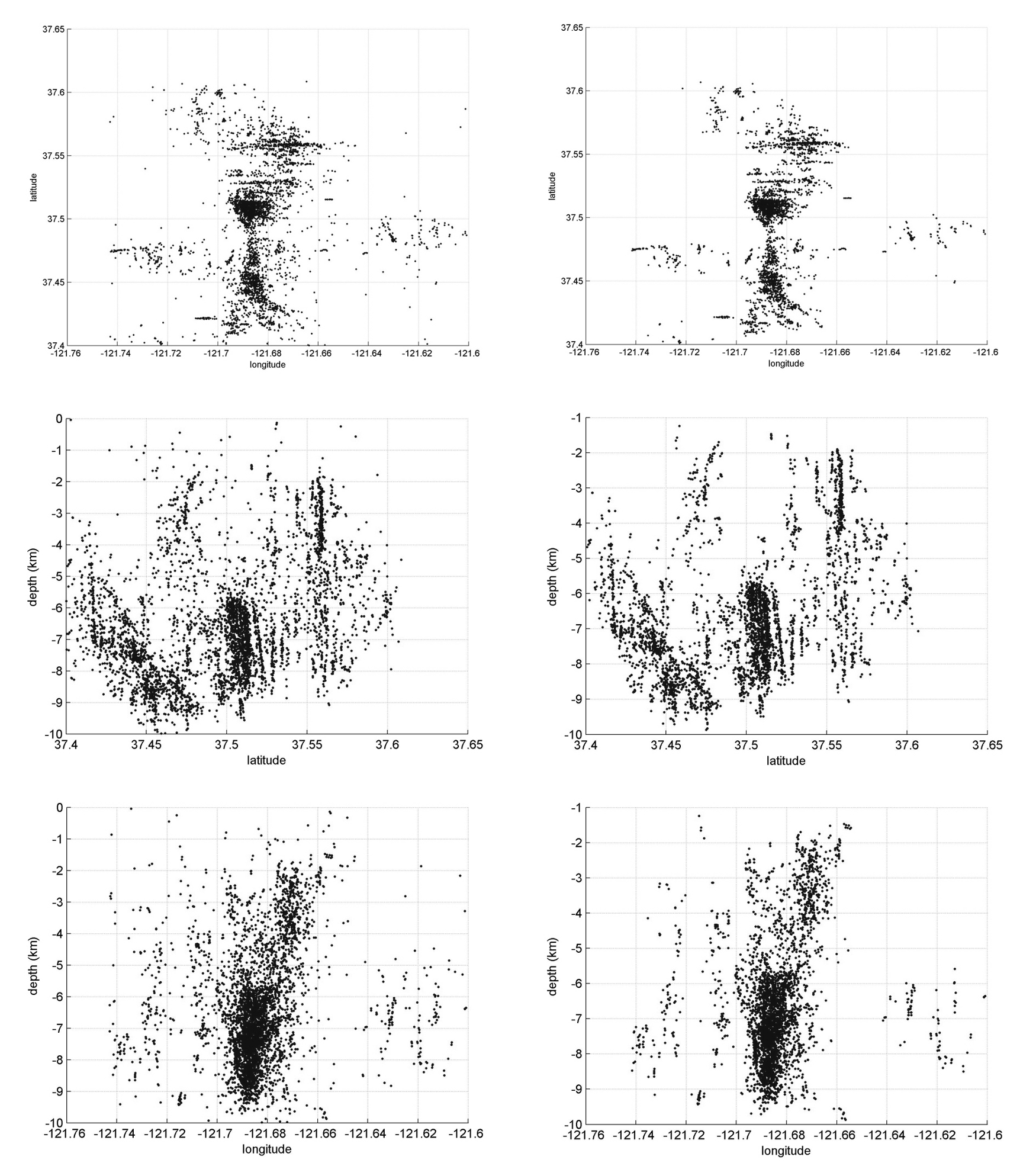}
\caption{\label{mntlewisseq} Spatial distribution of the seismic activity that followed the Mount Lewis event, 
shown along different projections. The left column shows the whole dataset; the right column shows the clustered part of seismicity (see text for details). }
\end{figure}

A caveat is that the definition of uncorrelated seismicity is time and space scale-dependent. For example, at large spatial scales, all events along the boundary between the Pacific and North American plates appear to be located within narrow and well-identified quasi-linear clusters. No part of the seismicity would be considered uncorrelated. But at smaller scales, as in  Fig. \ref{mntlewisseq}, many events appear quite distant from their nearest neighbors 
and do not define any specific pattern with them. Here, we avoid this issue of the scale dependence by focusing on the regional scale associated with the 1986 Mount Lewis earthquake. 

In contrast to the hypothesis underlying many previous studies and models, one should not assume that the uncorrelated seismicity is homogeneous. Following the concept that uncorrelated events belong to the subset of seismic activity occurring on under-sampled faults, its spatial organization should correspond to the under-sampling of a complex fault network. As an illustration of the distortions that may result from under-sampling, Eneva (1994) showed that an under-sampled monofractal distribution could be described by a spurious non-trivial multifractal spectrum.  

\begin{figure}
\noindent\includegraphics[width=15cm]{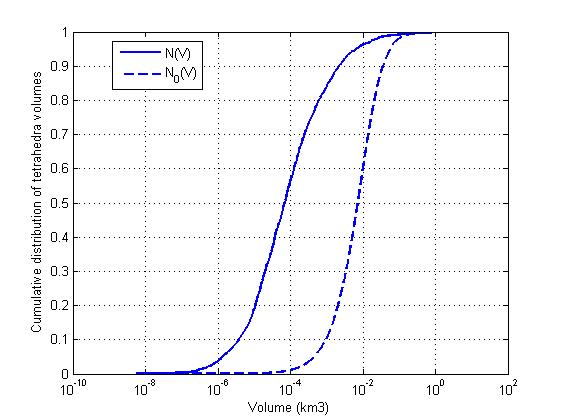}
\caption{\label{cumrealrand} Cumulative distribution $N(V)$ of the volume $V$ of tetrahedra formed with
four nearest neighbor events (see main text for details) for the full catalog (solid line) and the corresponding distribution $N_0(V)$ for randomized catalogs (dashed line).}
\end{figure}

We now proceed to identify the uncorrelated seismicity by comparing the distribution of natural events to a reference distribution of uncorrelated seismicity. Our method consists in comparing the cumulative distribution $N(V)$ of the volumes $V$ of tetrahedra formed with quadruplets of nearest neighbor events 
with distinct locations obtained from the natural catalog (Fig. \ref{cumrealrand} (continuous curve))  with the distribution $N_0(V)$ of the volumes of tetrahedra constructed with events belonging to a randomized catalog, where the horizontal (but not the vertical) dependence between events has
been destroyed (Fig. \ref{cumrealrand} (dashed curve)). We are guided by the logic that, in general, tetrahedra defined by clustered events should have 
smaller volumes than tetrahedra defined by unclustered seismicity events, because the former are supposed to be associated with local quasi-planar fault structures. The set of uncorrelated events is constructed by throwing at random (in the horizontal plane) the same number of events as in the real catalog, in a spatial domain whose size is identical to the natural one. Each event in the randomized catalog is located at the depth of a randomly chosen event in the real catalog.

Let $V_0(0.05)$ be the 5\%-quantile of the distribution $N_0(V)$ of tetrahedra volumes formed with events of the randomized  catalog, i.e., $N_0(V_0(0.05))=0.05$. We define the correlated seismicity as made of those events contributing to tetrahedra in the natural catalog with volumes smaller than $V_0(0.05)$.  Quantitatively, we find $V_0(0.05)=4.511 \times 10^{-4}$~km$^3$, which means that, by definition of the quantile at probability level $5\%$, only $5\%$ of tetrahedra volumes of the randomized catalog are smaller than $V_0(0.05)$. In contrast, we find  $N(V_0(0.05))=0.77$, which means that about $77\%$ of the tetrahedra constructed with events belonging to the natural catalog have a volume smaller than $V_0(0.05))$, thus reflecting a much stronger clustering on more planar structures. The difference between the natural and the randomized catalogs is also striking for the largest tetrahedra volume: $2$~km$^3$ for the former compared with $31$~km$^3$ for the latter.  

\begin{figure}
\noindent\includegraphics[width=16cm]{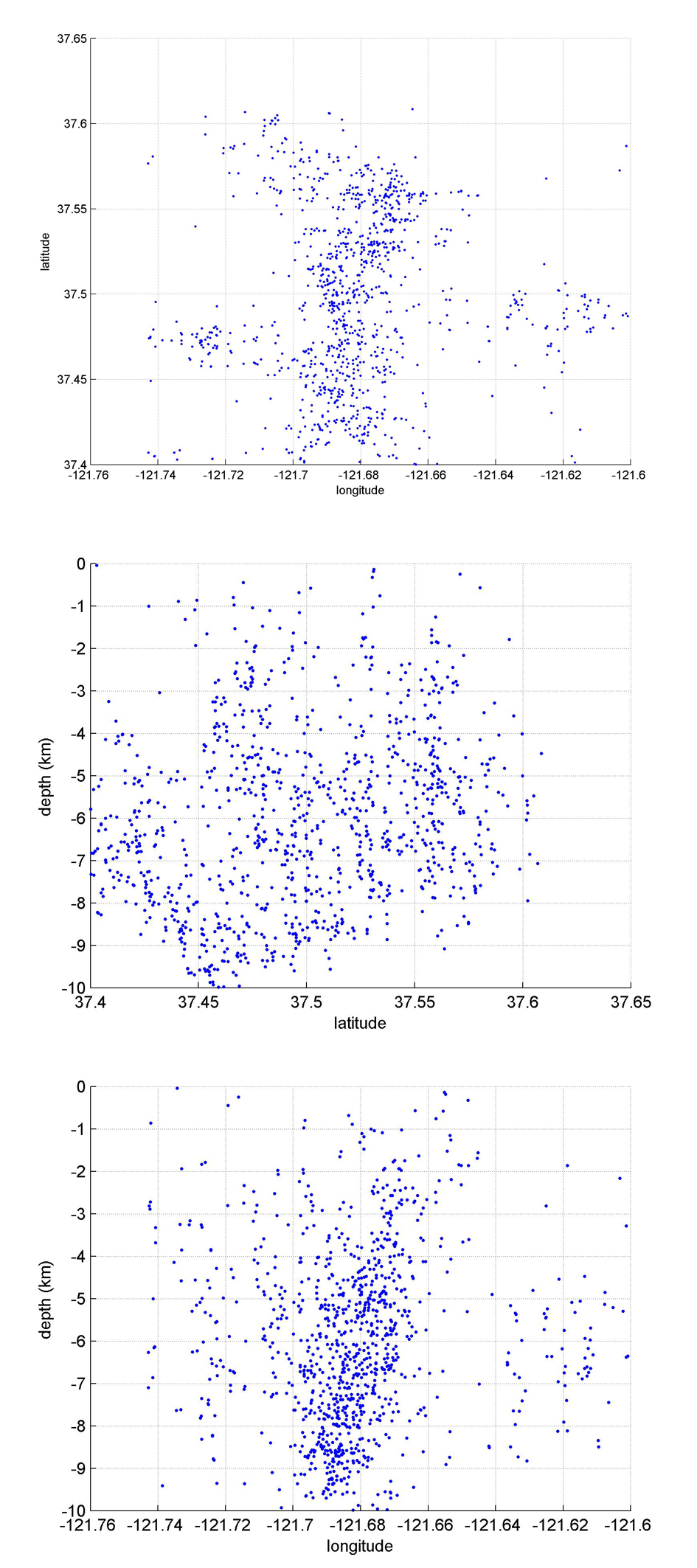}
\caption{\label{realunclust} Unclustered events detected in the catalog shown in Figure \ref{mntlewisseq}
using a local criterion using quadruplets of events forming tetrahedra compared between the initial and randomized catalogs (see text for details).}
\end{figure}

We thus select in the real catalog all events which belong to a tetrahedron of volume $V \le V_0(0.05)$. By construction, these events have only a $0.05$ probability of forming only by chance a tetrahedron with a small volume. All other events are removed and not considered in our analysis, as they are interpreted as belonging to under-sampled faults. This set of rejected events is shown in Figure \ref{realunclust}. Fig. \ref{mntlewisseq} (right column) shows three projections of the set of clustered seismicity. These three projections can be compared with those on the left column showing the full initial catalog, sum of both clustered and uncorrelated seismicity. In all three projections, the clustered part of seismicity appears to be much more strongly anisotropic than the full catalog. It is also very clear that the unclustered seismicity shown in Figure \ref{realunclust} does not exhibit clear patterns, in contrast with the clustered part shown in Fig. \ref{mntlewisseq} (right column). A local criterion using just quadruplets has been enough to sort out the seismicity of the aftershocks of 1986  Mount Lewis earthquake into two classes, which show very distinctive differences in clustering, already to the naked eye.

The tetrahedron is the simplest polyhedron, that allows us to define a volume, taken as a diagnostic of a non-planar pattern. This pre-treatment to sort out clustered and unclustered seismicity in terms of tetrahedra can be generalized to pentahedra and higher-order polyhedra, at the cost of more involved
calculations. Furthermore, our choice of the tetrahedron amounts to the minimal selection procedure, in the sense that the constraint is the most local. Using higher-order polyhedra would select higher-order statistical properties. This is not judicious for the goal of defining large scale fault-clusters, which by themselves should embody the information of the important large-scale coherent structures (the faults) of seismicity.

\subsection{Spatial data clustering of the Mount Lewis sequence}

We now apply our data clustering method to the clustered part of the Mount lewis sequence shown in Fig. \ref{mntlewisseq} (right column), as defined in the previous section. We shall first present the results of the cross-validation scheme, which allows to determine the number of kernels $K_{max}$ that should be used to fit the data, in a way similar to section \ref{appsynthex}. Then, we shall look at the obtained fault patterns using different values of $K$, before studying in more details the statistical properties of the clusters obtained for $K=K_{max}$, i.e. of the optimal set of fault planes proposed to explain the earthquake catalog.

\subsubsection{Cross validation results}

\begin{figure}
\noindent\includegraphics[width=15cm]{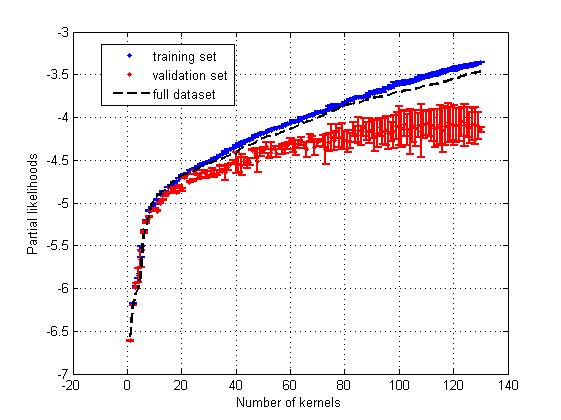}
\caption{\label{lewis_crossvalid} $L_{pd}$s of the training dataset (upper curve) and of the validation dataset (lower curve) for the Mnt Lewis catalogue with $p=0.1$. The middle dashed curve shows the $L_{pd}$ of the fit of the full catalogue ($p=0$).}
\end{figure}

Figure \ref{lewis_crossvalid} shows the dependence on $K$ of the $L_{pd}$'s of the training and validation datasets (using $p=0.1$). This figure is thus analog to Fig. \ref{synth_crossvalid}. As observed for the synthetic dataset in section \ref{appsynthex}, error bars are much larger for the validation dataset (lower curve) than for the training dataset (upper curve). The middle dashed curve shows, for comparison, the dependence of $L_{pd}$ when we fit the full dataset (so that $p=0$ ; there is thus no validation dataset in this case). The $L_{pd}$ of the training dataset in the case $p=0.1$ is larger than the $L_{pd}$ when using the full dataset because the former features less events, which makes the fit easier. This Figure shows that the $L_{pd}$'s increase very quickly with $K$ when less than $10$ kernels are used for the fit. Then, the likelihoods increase more slowly without exhibiting any bound for the training sets, as increasing $K$ provides better fits. In the case of the validation set, the increase is even slower, and the $L_{pd}$'d saturate at a constant value when $K>100$, suggesting that increasing $K$ further does not bring any significant information about the structure of the dataset. However, unlike the synthetic case, the $L_{pd}$ of the validation set does not decrease when $K$ increases beyond a critical value, which makes an accurate measure of $K_{max}$ much more difficult. This figure displays results obtained for $K=1$ to $130$, suggesting that $K_{max} \simeq 100$. Fitting the training dataset with $K$-values larger than $130$ proved very unstable as we found an increasing number of cases where Gaussian kernels collapse on clusters featuring only $3$ or less events, preventing the convergence of the algorithm converges. Note that, in the case where such a collapse is found, the corresponding kernel is removed and we re-start a fit with $K-1$ kernels.

\begin{figure}
\noindent\includegraphics[width=15cm]{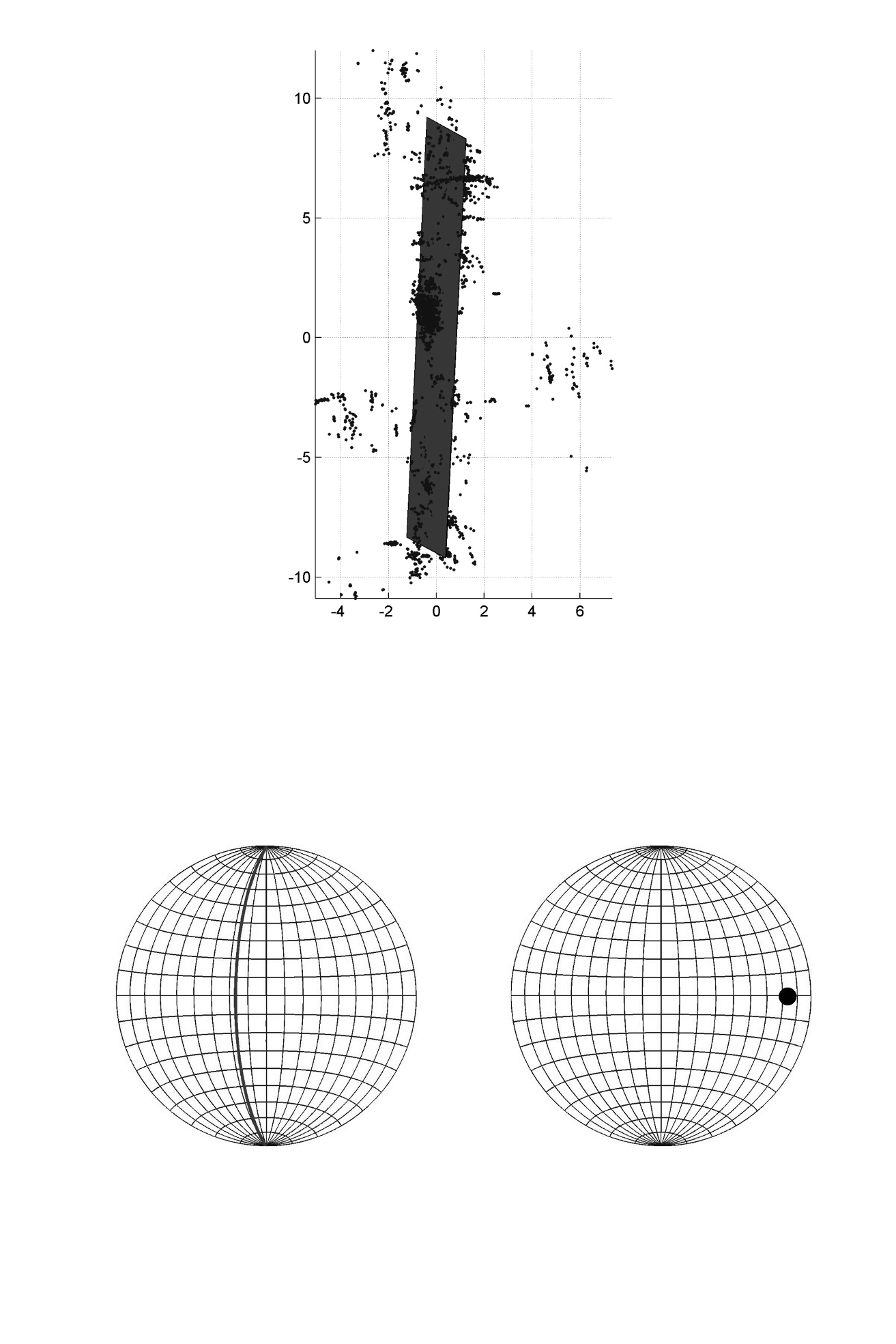}
\caption{\label{lewis_clean_1_plane} Fit of the catalog of clustered events shown in Figure \ref{mntlewisseq} with 1 plane (top) and its associated stereographical projections (bottom). All scales in $km$.}
\end{figure}

\begin{figure}
\noindent\includegraphics[width=15cm]{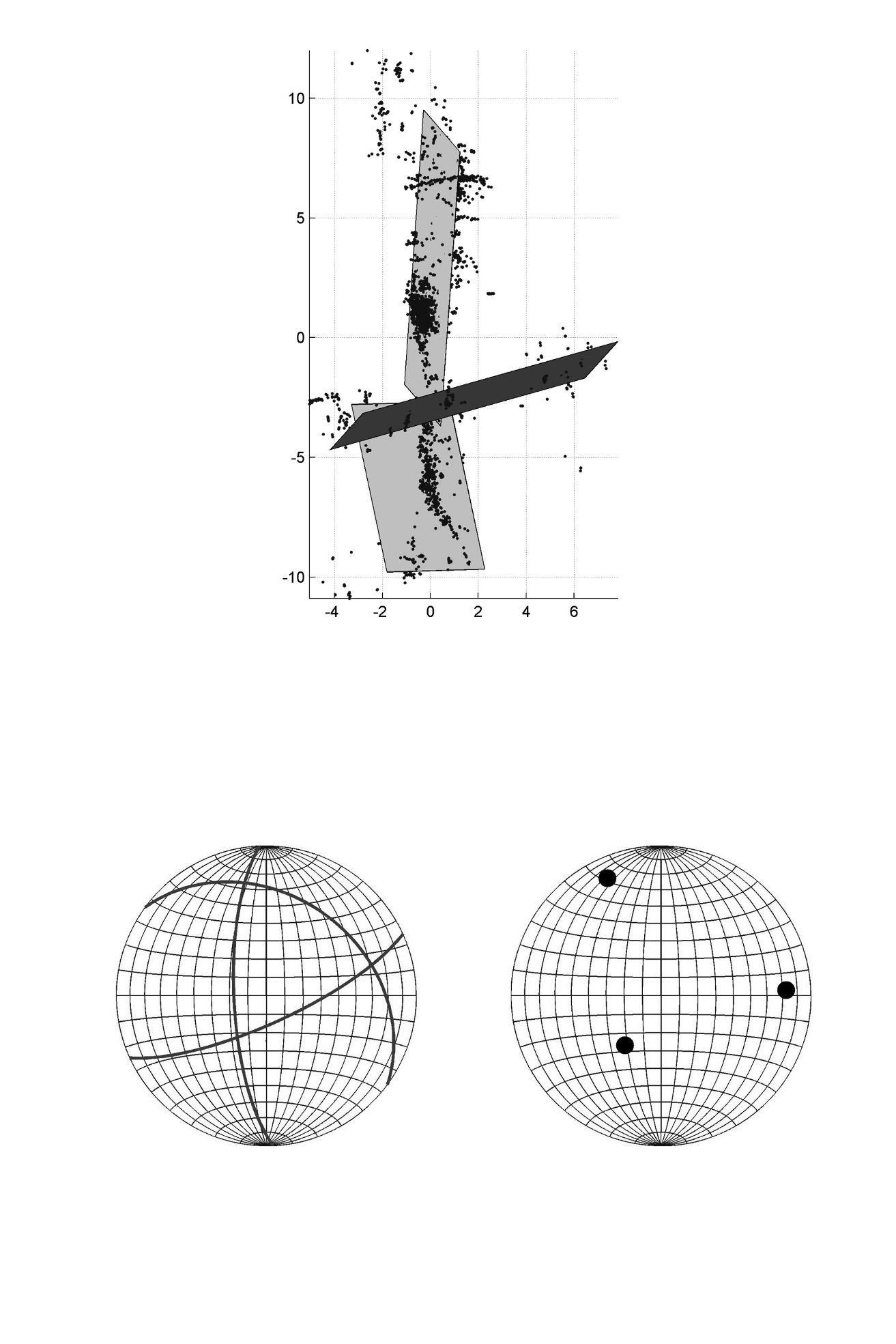}
\caption{\label{lewis_clean_3_plane} Fit of the catalog of clustered events shown in Figure \ref{mntlewisseq} with 3 planes (top) and their associated stereographical projections (bottom). All scales in $km$.}
\end{figure}

\begin{figure}
\noindent\includegraphics[width=15cm]{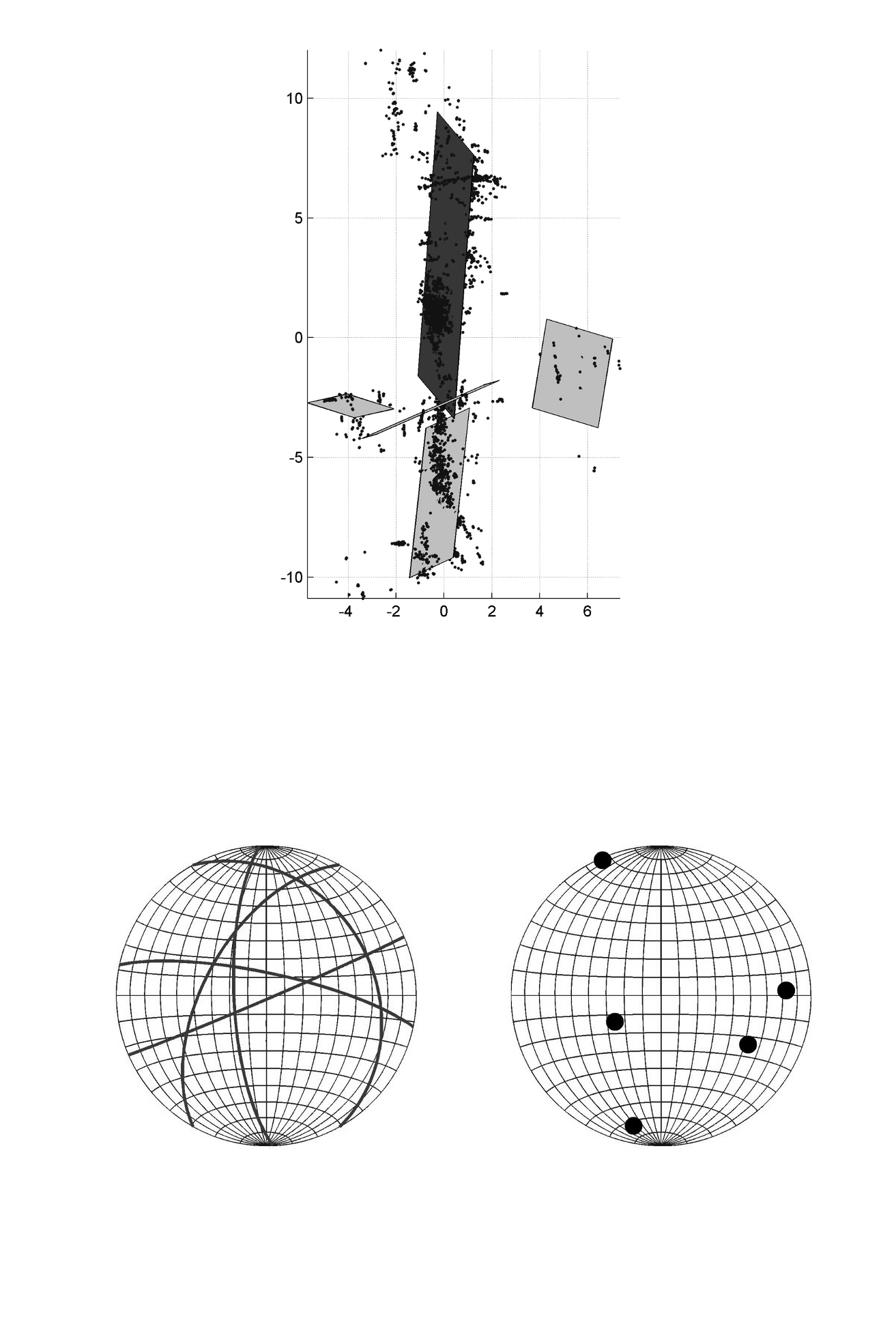}
\caption{\label{lewis_clean_5_plane} Fit of the catalog of clustered events shown in Figure \ref{mntlewisseq} with 5 planes (top) and their associated stereographical projections (bottom). All scales in $km$.}
\end{figure}

\begin{figure}
\noindent\includegraphics[width=15cm]{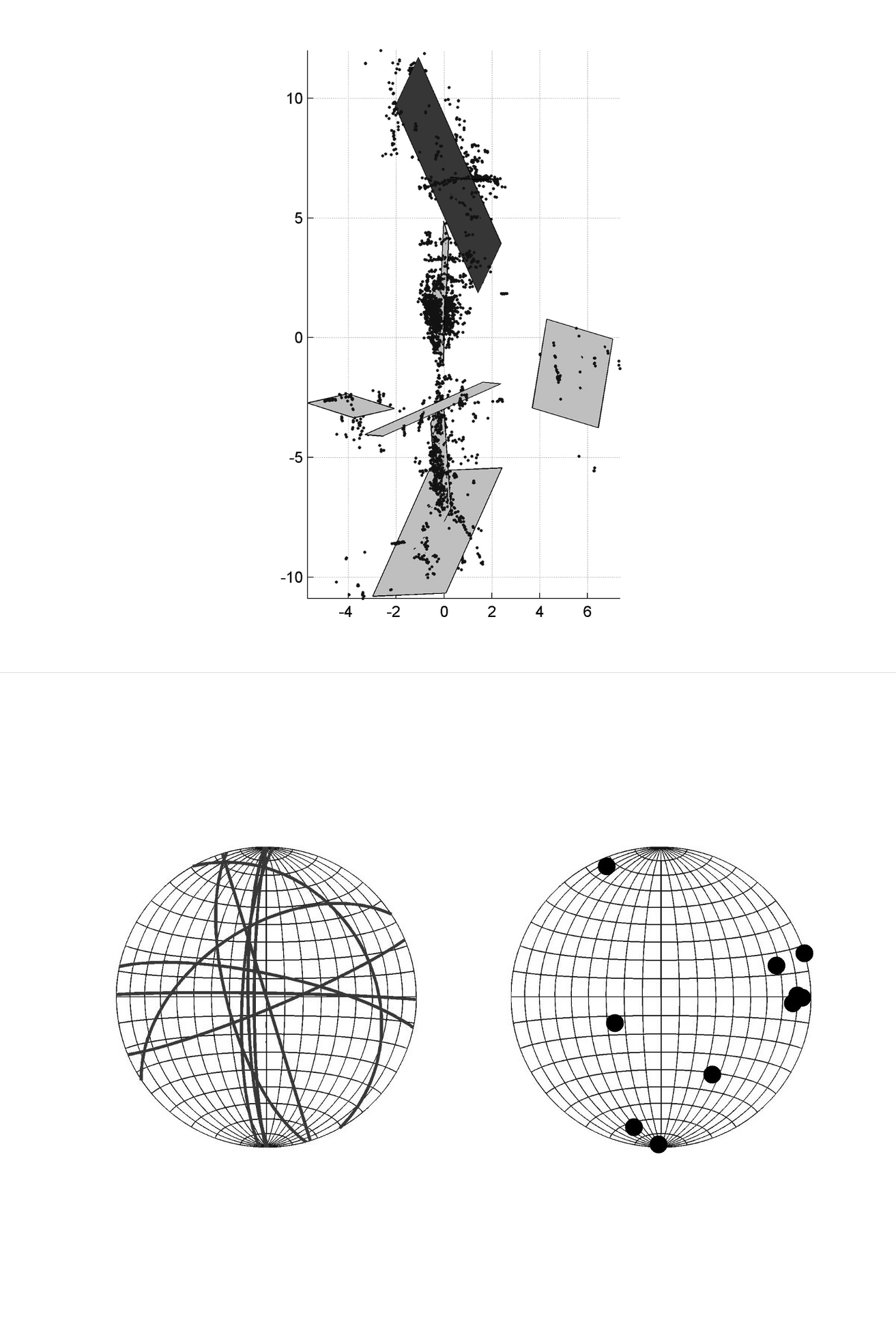}
\caption{\label{lewis_clean_10_plane} Fit of the catalog of clustered events shown in Figure \ref{mntlewisseq} with 10 planes (top) and their associated stereographical projections (bottom). All scales in $km$.}
\end{figure}

\begin{figure}
\noindent\includegraphics[width=15cm]{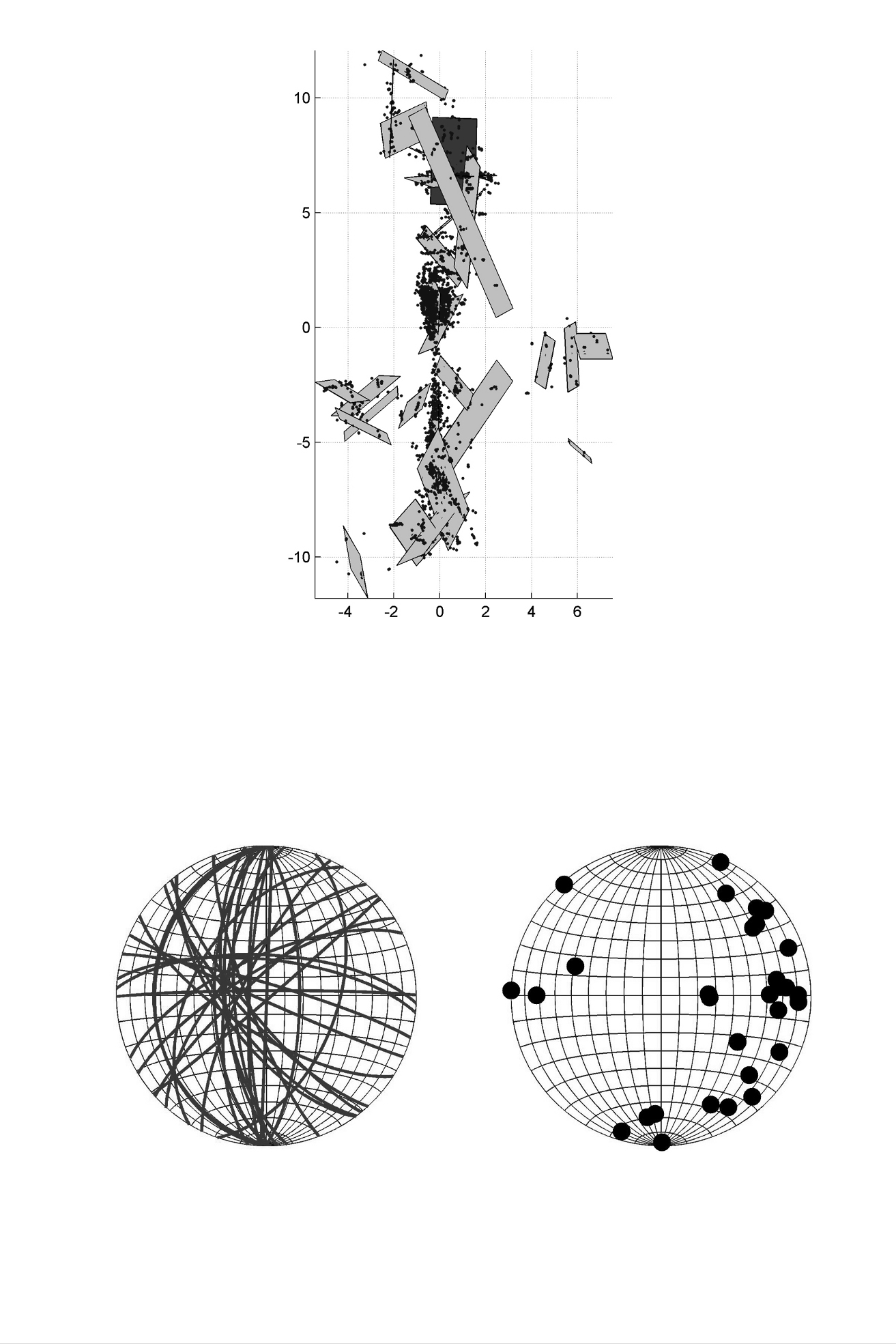}
\caption{\label{lewis_clean_30_plane} Fit of the catalog of clustered events shown in Figure \ref{mntlewisseq} with 30 planes (top) and their associated stereographical projections (bottom). All scales in $km$.}
\end{figure}

\begin{figure}
\noindent\includegraphics[width=15cm]{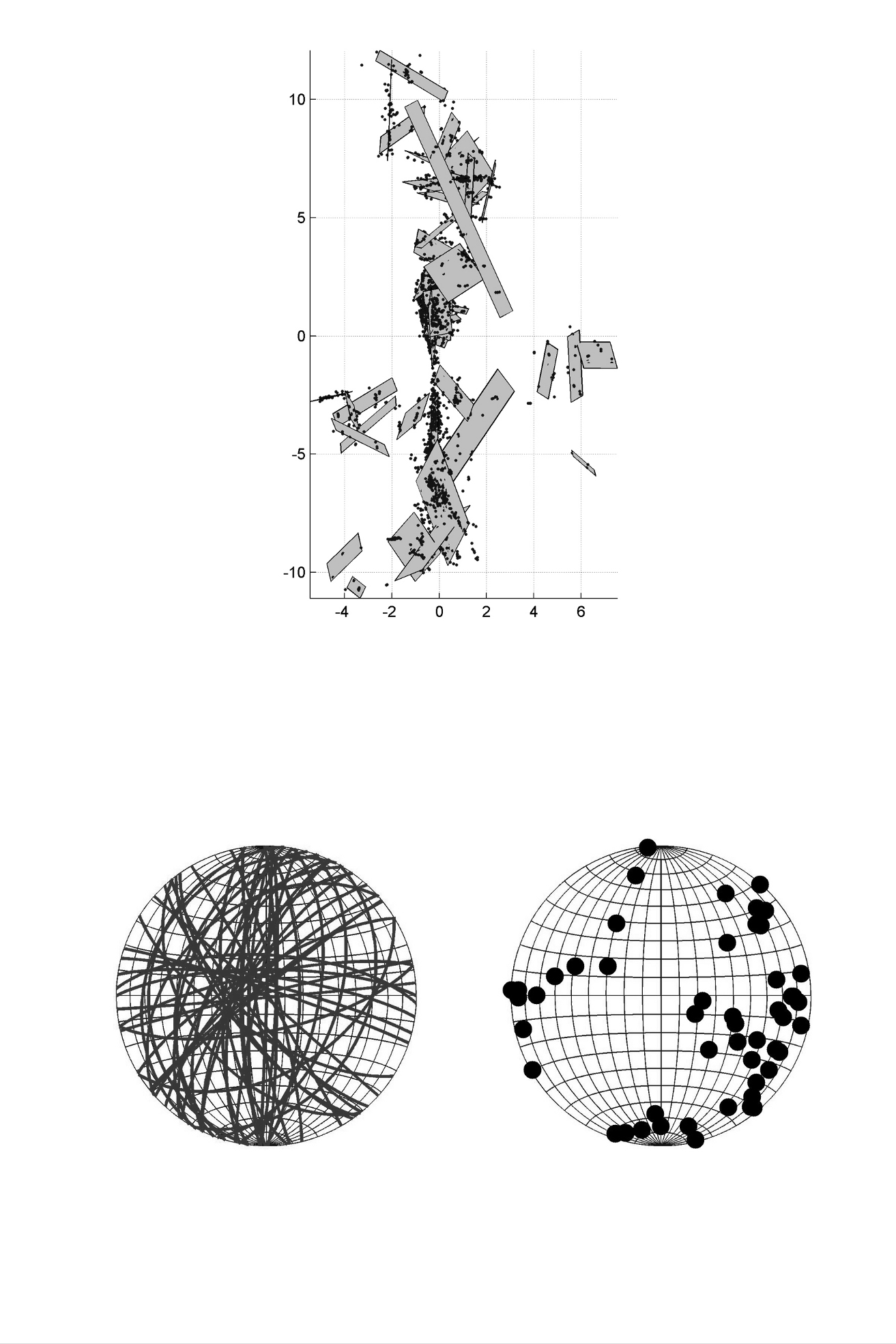}
\caption{\label{lewis_clean_50_plane} Fit of the catalog of clustered events shown in Figure \ref{mntlewisseq} with 50 planes (top) and their associated stereographical projections (bottom). All scales in $km$.}
\end{figure}

\begin{figure}
\noindent\includegraphics[width=15cm]{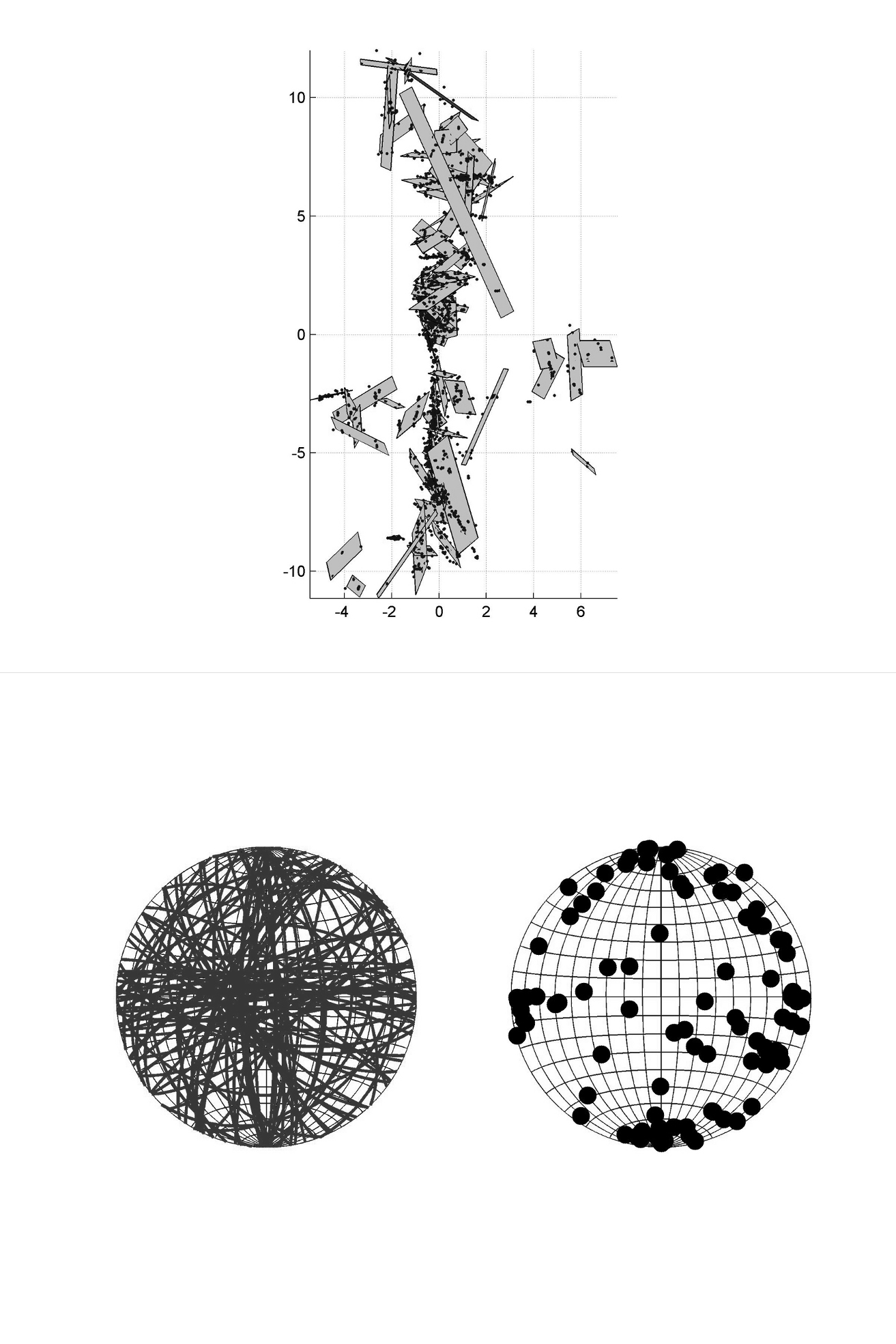}
\caption{\label{lewis_clean_90_plane} Fit of the catalog of clustered events shown in Figure \ref{mntlewisseq} with 90 planes (top) and their associated stereographical projections (bottom). All scales in $km$.}
\end{figure}

\begin{figure}
\noindent\includegraphics[width=15cm]{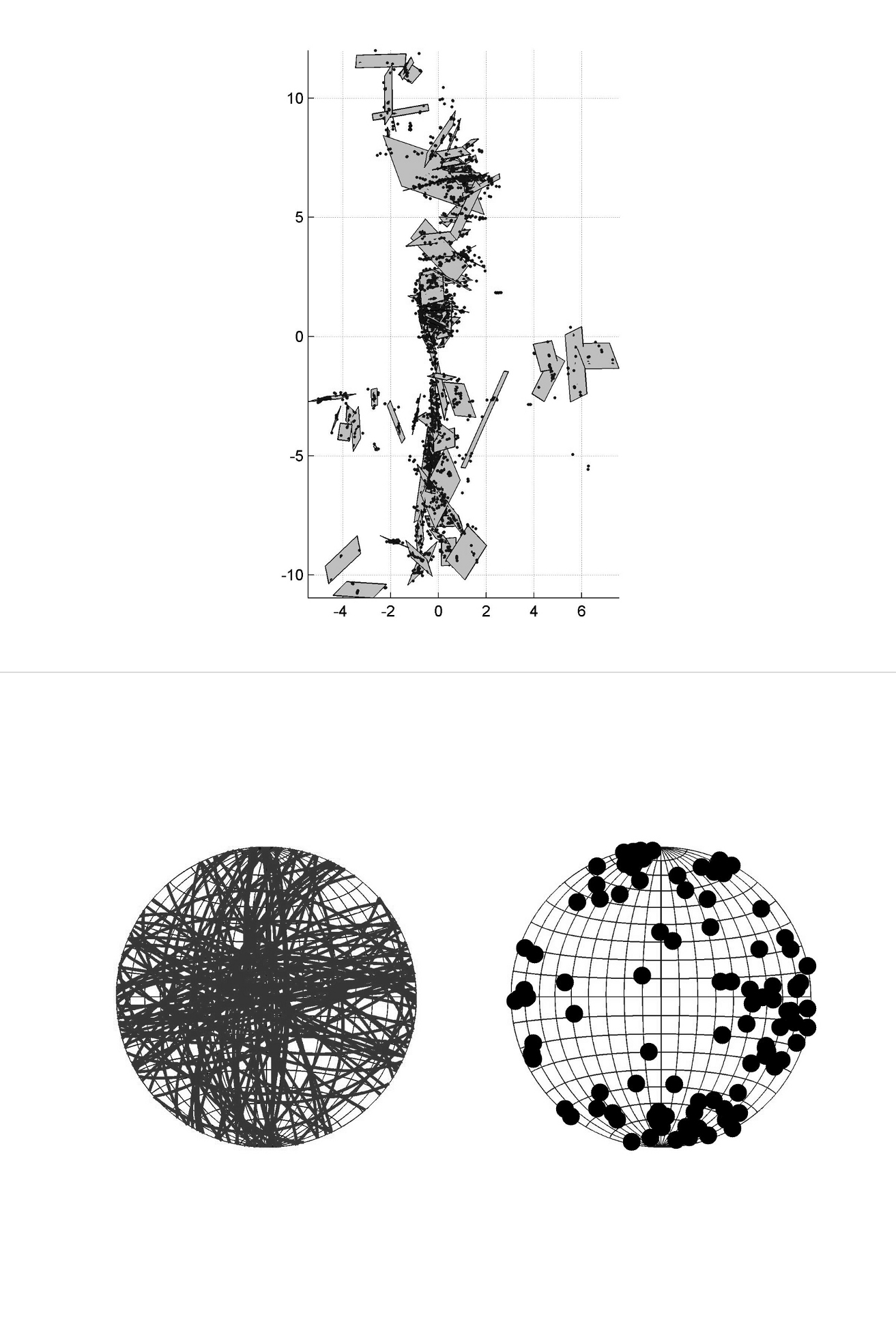}
\caption{\label{lewis_clean_100_plane} Fit of the catalog of clustered events shown in Figure \ref{mntlewisseq} with 100 planes (top) and their associated stereographical projections (bottom). All scales in $km$.}
\end{figure}

\begin{figure}
\noindent\includegraphics[width=15cm]{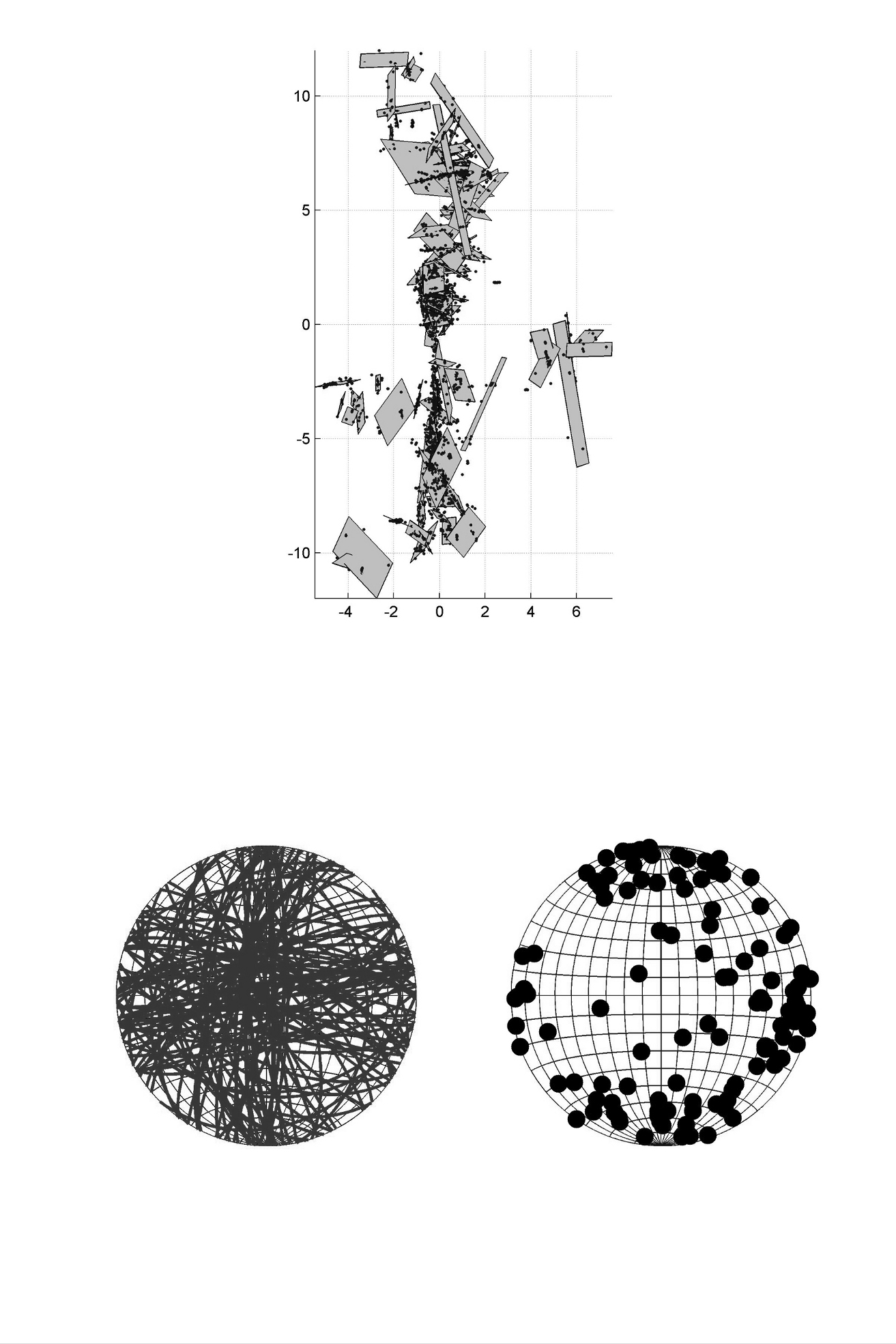}
\caption{\label{lewis_clean_110_plane} Fit of the catalog of clustered events shown in Figure \ref{mntlewisseq} with 110 planes (top) and their associated stereographical projections (bottom). All scales in $km$.}
\end{figure}

\begin{figure}
\noindent\includegraphics[width=15cm]{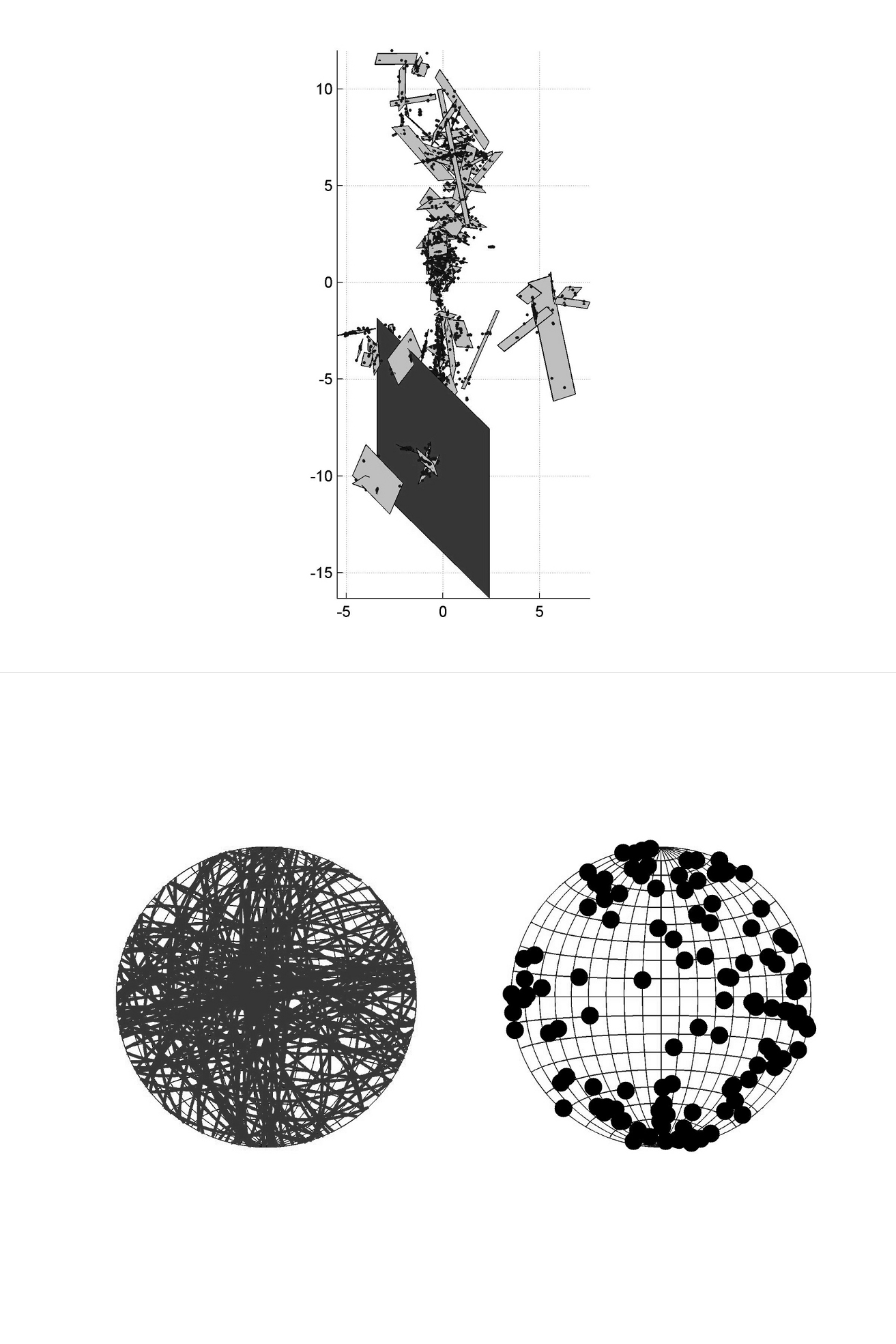}
\caption{\label{lewis_clean_120_plane} Fit of the catalog of clustered events shown in Figure \ref{mntlewisseq} with 120 planes (top) and their associated stereographical projections (bottom). All scales in $km$.}
\end{figure}

\begin{figure}
\noindent\includegraphics[width=15cm]{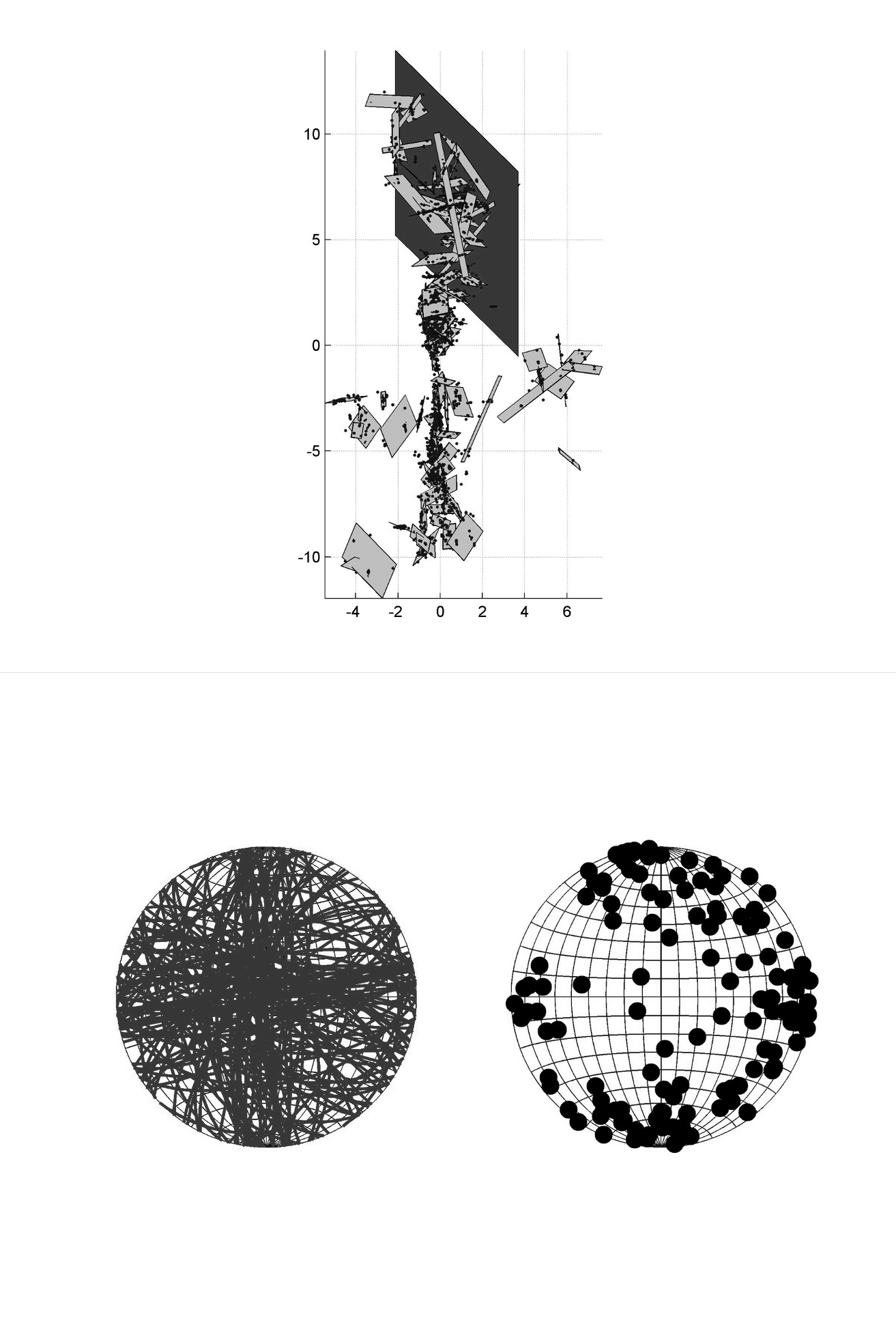}
\caption{\label{lewis_clean_130_plane} Fit of the catalog of clustered events shown in Figure \ref{mntlewisseq} with 130 planes (top) and their associated stereographical projections (bottom). All scales in $km$.}
\end{figure}

We now switch to the results obtained when fitting the whole dataset with a varying number of clusters. For example, Figures \ref{lewis_clean_1_plane} to \ref{lewis_clean_10_plane} show the set of fitting planes obtained when increasing $K$ from $1$ to $10$ (we only show the cases $K=1, 3, 5, 10$). Figures \ref{lewis_clean_30_plane} to \ref{lewis_clean_130_plane} show the obtained solutions when increasing $K$ from $20$ to $130$ (we only show the cases $K= 30, 50, 70, 90, 100, 110, 120, 130$). In each figure, the upper plot shows a $2D$ epicentral projection of the dataset as well as of the set of planes. The darkest plane is the one that will be split in the next step by going from $K$ to $K+1$ to form the initial conditions of the EM procedure. The lower plot of each figure shows a stereographic projection of the corresponding set of planes (on the left) and of the associated set of poles (on the right). The plots on the left show, as seen from above, the intersection of the planes with the lower hemisphere of a unit sphere, assuming that all planes have been translated in space so that their centers coincide with that of the sphere. It thus provide a direct visualization of the anisotropy of the set of fault traces. The plots on the right show the intersection with the same hemisphere of the set of vectors normal to each plane (one extremity of the vector being attached to the center of the unit sphere). This plot is particularly useful when many faults are plotted. Both plots are indeed equivalent and help to visualize quickly the main anisotropy directions if any. For example, in the case of poles, vertical planes correspond to points located on the edge (equator) of the plot, while horizontal planes correspond to points located at the center of the plot. 
All stereographic plots indicate that a large majority of clusters define nearly vertical structures, as most poles concentrate near the equator, yet their azimuths are highly variable. It also seems that a majority of clusters dip to the West, whereas the focal mechanism of the main rupture indicated a dip to the East (note that the main shock belongs to the set of unclustered events and is thus not part of the fitted dataset). We can observe that, for a number of clusters larger than sixty, more and more planes dip to the East. We also notice that most of the newly added planes are then generated in the central part the catalog (within $[0 ; 3]$~km along the vertical direction), suggesting that this small-scale area features a more or less disorganized structure. If this is the case, the associated set of events would reveal the fuzzy structure of a fault-zone (as depicted on Fig. \ref{faultsismo}).

Figures \ref{lewis_clean_120_plane} and \ref{lewis_clean_130_plane} both display a very large anomalous cluster (which is about to be splitted in the next iteration), with a thickness of several kilometers. This solution does not show up systematically for very large values of $K$ (for example, it is not observed when $K=123$), but seems to be more frequent when $K$ is significantly larger than $100$. Our interpretation is that clusters are first used to fit small scale features (which correspond to the largest fluctuations of 
the density of local events) to maximize the likelihood; then, when all small-scale features are correctly fitted, supplementary kernels (i.e. for $K > 100$) are possibly used to fit larger scale residuals of the distribution, hence the apparition of such anomalous large scale clusters.

Fig. \ref{lewis_crossvalid} shows that the error bars preclude the determination of a very precise value of $K_{max}$. 
Once a value of $K_{max}$ has been chosen, we can fit the whole dataset using that optimal number of kernels. 
In the following, we shall thus discuss and compare in more details the statistical properties of individual clusters as a function of $K$, focusing on their orientation and size. We shall see that the results also suggest $K_{max} \simeq 100$.

\subsubsection{Anisotropy properties}

The standard way for studying the anisotropy properties of a given fault network is the use of stereographical plots. These plots require the knowledge of the strike $\lambda$ and the dip $\delta$ of the clusters. We now introduce another parameter $\beta$ which we simply define as $\beta = \lambda \bmod \pi$, which is thus  a measure of the azimuth of a fault, independently of its dip.

\begin{figure}
\noindent\includegraphics[width=15cm]{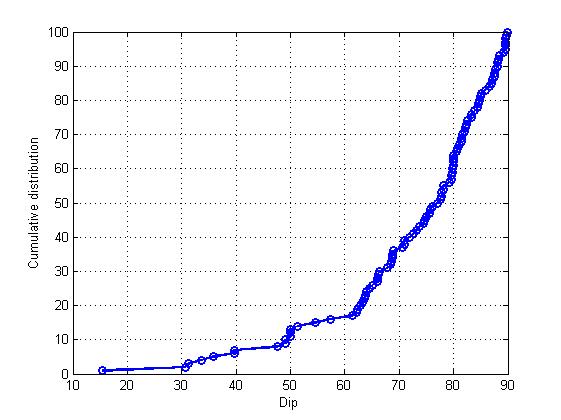}
\caption{\label{lewis_100_dipcum} Cumulative distribution of fault dips for the fit of the Mnt Lewis sequence with $K=100$ kernels.}
\end{figure}

The distribution of dips does not show a strong dependence on $K$ when $K>100$. When binning the dip axis by steps of $10^{\circ}$, we find that the entropy of the histogram of fault dips reaches a local approximate minimum for $K \simeq 100$. Performing the same analysis on the histogram of $\beta$ values, a minimum entropy is also observed when $K \simeq 100$. This suggests that, for $K<100$, using too few kernels to fit the data forces some of them to gather small-scale uncorrelated clusters into larger ones with a more or less random orientation. When $K>100$, we try to explain the data with too many kernels, so that we fit smaller scale features that are likely to possess a geometry depending more heavily on location errors: they are thus oriented more randomly too. This result once again suggests that the optimal number of clusters that one should use to fit the catalog should be set to $K \simeq 100$. When considering the joint distribution of $(\lambda, \delta)$ (by binning both axes by steps of $10^{\circ}$), the corresponding entropy monotonically increases with $K$, which tempers the previous conclusion. The minimum observed dip for the fits obtained for $K=100$ to $K=130$ varies between $13^{\circ}$ and $17^{\circ}$, but the solution for $K=100$ features the largest proportion of planes with a dip larger than $30^{\circ}$ ($99\%$), as well as the largest proportion ($84\%$) of planes with a dip larger than $60^{\circ}$. The cumulative distribution of dips for $K=100$ is shown on Fig. \ref{lewis_100_dipcum}. 

\begin{figure}
\noindent\includegraphics[width=15cm]{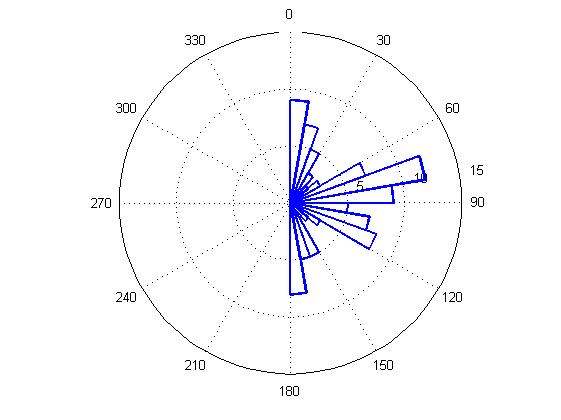}
\caption{\label{lewis_100_rose_mod_pi} Histogram of $\beta$ (azimuth of fault trace) for $K=100$.}
\end{figure}

Fig. \ref{lewis_100_rose_mod_pi} shows a rose histogram of the observed $\beta$ values in the case $K=100$. One can clearly notice three main directions of faulting: the main one strikes about NS, while the other ones strike respectively $N75$ and $N110-115$ (and thus seem to be conjugate directions of faulting). No correlation was found between dip and strike or $\beta$. We shall now look at the distribution of the size of clusters.

\subsubsection{Size and shape of clusters \label{mntlewissize}}

\begin{figure}
\noindent\includegraphics[width=15cm]{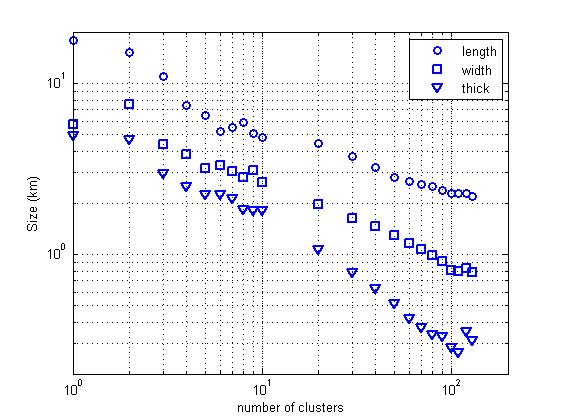}
\caption{\label{lewis_size_mean} Mean length, width and size of clusters as a function of $K$.}
\end{figure}

Figure \ref{lewis_size_mean} shows the mean size of clusters (length $L$, width $W$ and thickness $T$) as a function of $K$. The length and width of clusters have already been defined in section \ref{appsynthex}. The thickness is defined as $4\sigma_3$, where $\sigma_3$ is the standard deviation of events in a direction normal to the fault plane (the factor $4$ ensures that the thickness corresponds to the spatial domain covered by $96\%$ of the associated events when the cluster obeys a genuine Gaussian distribution). The volume of a cluster is simply defined as the product of the three former quantities.

Fig. \ref{lewis_size_mean} reveals several features: first, all quantities decrease with $K$ up to $K=100$, and seem to reach a constant value beyond this threshold. The decay stems from the ability to fit smaller features by adding more kernels, but values of $K$ larger than $100$ seem of no use to decipher the small scale structure of the catalog. This is thus a supplementary argument to propose $K \simeq 100$ as the maximum number of kernels to use to fit this dataset. The linearity of the log-log plots in Fig. \ref{lewis_size_mean} for large values of $K$ (from $20$ up to $100$) reveals underlying power-laws: $L \sim K^{-0.4}$, $W \sim K^{-0.55}$ and $T \sim K^{-0.85}$. These power laws can be restated as

\begin{equation}
K \sim L^{-2.5}~, \quad  W \sim L^{1.37}~,  \quad  T \sim L^{2.1}~.
\label{hyybwqa}
\end{equation}

The first scaling law  $K \sim L^{-2.5}$ may correspond to a measure of the capacity fractal dimension 
$D_f  =2.5$ of the reconstructed fault network. The two other scaling laws express the self-affinity
properties of the faults in the network. The fact that the exponents $0.55/0.4=1.37$ and $0.85/0.4= 2.1$
are larger than $1$ implies that the scaling laws hold in the small size asymptotic (so-called
``ultraviolet'' regime): they suggest that small fault segments tend to become more and more one-dimensional
with very thin width, the smaller they are.  This is in agreement with the mechanics
of shear localization, in which the smallest structures are localized at the scale of grain sizes
while large structures are processed zones with thicker cores.

Relating the number $K$ of clusters to the mean volume of the clusters (defined as the product of their length, width and thickness)
provides an estimate of the average properties of the fault network with respect to a unique size measure.
Considering a characteristic length $\epsilon$ equal to the cubic root of the mean volume, we plot $K$ as a function of $\epsilon$ in Fig. \ref{lewis_fractal}. The two anomalous data points on the left of the plot correspond to $K=120$ and $K=130$, and we already discussed their origin. Discarding these two points, the log-log plot appears to be linear for small values of $\epsilon$, showing that $K \simeq \epsilon^{-D}$ with $D \simeq 1.85$. This exponent compares well with the fractal dimension $D_0=1.8-2.0$ of hypocenters in Southern California proposed by Robertson et al (1995) using a high quality relocated catalog. Yet, we must ackowledge that the scaling we observe holds over a very limited range of scale, due to the small spatial extent of the dataset.

\begin{figure}
\noindent\includegraphics[width=15cm]{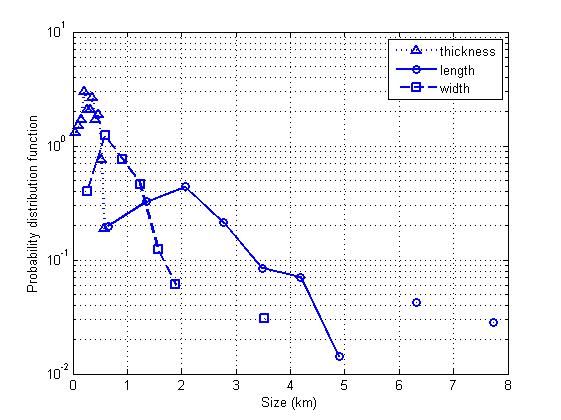}
\caption{\label{lewis_size_pdf_100} Probability distribution of length, width and thickness of clusters for $K=100$.}
\end{figure}

Figure \ref{lewis_size_pdf_100} shows the probability density function of the length, width and thickness of clusters for $K=100$. All distributions span a very limited scale interval, so that inferring the theoretical underlying distribution is not obvious. However, for length and width, this semi-log plot shows that both are compatible with exponential distributions. The exponential decay suggests characteristic sizes of $\simeq 2.4$~km for length and $\simeq 0.9$~km for width. Both values are close from the true averages of cluster length and width for $K=100$ (respectively $2.27$~km and $0.81$~km). The thickness of clusters spans a much smaller scale interval and does not seem to obey any simple analytical form. It is almost flat in-between its extreme values. It is worth noticing that, when $K=100$, the average value of $\sigma_3$ is about $72$~m, while the average uncertainty of relative locations is about $8$m (one standard deviation). This suggests that the observed thickness of clusters does not stem from location uncertainties, but is rather a genuine geometrical feature of fault zones.

\begin{figure}
\noindent\includegraphics[width=15cm]{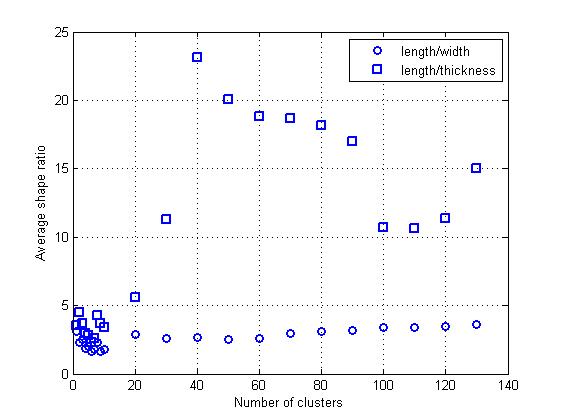}
\caption{\label{lewis_ratios} Dependence of mean shape ratios $L/W$ and $L/T$ as a function of $K$.}
\end{figure}

\begin{figure}
\noindent\includegraphics[width=15cm]{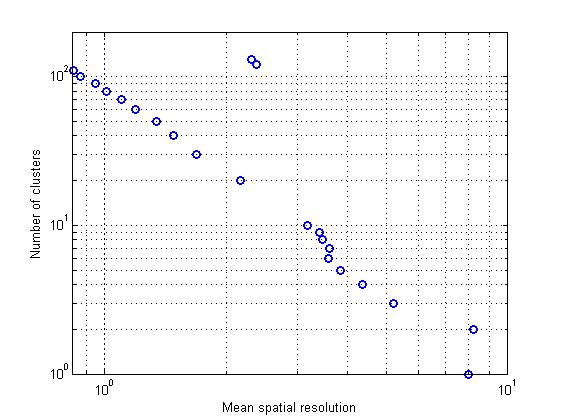}
\caption{\label{lewis_fractal} Relationship between the average spatial resolution $\epsilon$ and $K$.}
\end{figure}

We were unable to identify any correlation between the size parameters of clusters (length, width or thickness) for given values of $K$. This may be due to the very limited interval covered by those sizes (about one order of magnitude at most). We thus computed the average value of the ratios $L/W$ (hereafter $\rho_{LW}$) and $L/T$ (hereafter $\rho_{LT}$) over all clusters for fixed values of $K$. Figure \ref{lewis_ratios} shows that the ratio $\rho_{LW}$ is approximately constant with $K$, suggesting that, at any scale, the length of clusters is about $3$ times their width (for $K=100$ we get $\rho_{LW}=3.37$). The other shape ratio, $\rho_{LT}$, shows much wilder fluctuations with $K$, and is about $10.7$ for $K=100$. Clusters thus appear as quite elongated structures with significant anisotropy in all spatial directions. Note that those results differ a little from what can be deduced from Eq. \ref{hyybwqa}, as the latter allows to define ratios of the means and not means of the ratios.


\section{Discussion and Conclusion \label{conclusion}}

A fundamental and still unresolved question in brittle tectonics deals with the mechanics of fault growth. For example, strike-slip faults generally display discrete linear fault segments offset by compressive or tensile fault steps. It is believed that, as time increases, those steps (or any other non-planar fluctuation) are progressively damaged and destroyed to allow segments linkage into larger scale segments, themselves offset by larger scale steps (De Joussineau and Aydin, 2009; Ben Zion and Sammis, 2003). It seems  that this process is self-affine (and not self similar as claimed in De Joussineau and Aydin (2009)) so that the fault appears smoother and smoother as cumulative slip increases. In the end, combined with the mechanics of constrained fragmentation in the fault gouge, the active part of the fault reduces to a very well defined principal fracture surface which is nearly an euclidean surface (Ben Zion and Sammis, 2003). In that picture, the damage zone around the principal fracture surface that is observed in the field is just a record of the past fault history and does not host any significant seismic activity. If surface field observations seem to support that model, it is still a matter of debate for faulting at seismogenic depth, notwithstanding the fact that this view may hold for mature faults but might be only a rough approximation for younger faults. On other grounds, fault segmentation is an important feature of faults that ultimately controls the initiation, size and dynamics of co-seismic ruptures, while the damaged zone may play a significant role in crustal fluid flow or act as a guide for surface waves.

Fault segmentation was up to now largely inferred from the statistical analysis of fault maps. A few geophysical techniques may also be used for imaging tectonic structures at depth, but their rather poor resolution precludes
segregating active from inactive components of the fault structures. In this work, we have proposed a new approach which takes account of seismicity itself to infer the three dimensional dynamics of faulting at depth. We are aware that earthquake catalogs do not sample properly and uniformly the deformation processes at depth (for example we miss aseismic creep events and slow earthquakes), but they reveal at least the minimum spatial extent of active processes. Earthquakes are never distributed in a uniform manner but display strong  and complex clustering properties, which may even change with the considered time scales. The spatial shape of those clusters may nevertheless bear similarities with the structure of faults. We thus used a data clustering approach (based on the expectation-maximization algorithm) to provide a probabilistic reconstruction of spatial clusters within an earthquake catalog, without taking account of their causal relationships. The main idea we developed is to approximate the local density of events by a linear superimposition (a mixture) of Gaussian kernels. This particular shape of kernels ensures that we do not assume anything about the clusters, except that they can be characterized by a position, an orientation and a finite spatial extent. This mixture approach, when combined with a cross validation scheme, allows to determine the number of clusters that compose the analyzed catalog, i.e., it provides a maximum bound for the number of kernels needed to approximate coherent structures in the catalog. In 
the case of the Mount Lewis sequence that we have analyzed, we found that $\simeq 100$ clusters are
needed to explain the sequence of events that followed the Mount Lewis, 1986 event. 

The properties of those clusters have been analyzed using standard statistical techniques. We could observe that the orientation of clusters is not random but reveals the existence of three distinct directions of faulting. The main one (nearly $N0$) corresponds to the current boundary conditions in Northern California and is thus correlated with younger structures. It is for example fully compatible with the direction of the fault plane given by the focal mechanism of the main event. The two other directions ($N70$ and $N110$) are most likely inherited from previous tectonic history. Our results show that those older structures still play a significant role in the seismogenesis in that area, and thus should not be discarded in seismic hazard analyses. 

The spatial extent of clusters provides an important insight about the scale of fault segmentation, which is thought to be a function of the cumulative slip over a fault (De Joussineau and Aydin, 2009). We showed that both the length and width of clusters are exponentially distributed, i.e. display characteristic values. The typical length of clusters is found about $2-2.5$~km, while their typical width is about $0.8-0.9$~km. Such high resolution can only be achieved when using high-quality catalogs built with modern techniques of earthquake location. Assuming that the clusters we identified is made of fault segments, then Fig. $3$ of De Joussineau and Aydin predicts that the maximum slip on the underlying faults should lie within $[200 {\rm m} ; 2 {\rm km}]$. It is worth noticing that the size of the clusters is comparable to the assumed size of the main shock rupture, so that the latter may have broken a single fault segment. It is also similar to the typical fault segment length empirically observed by Kilb and Rubin (2002) in the same area.

The last typical size of clusters is their thickness, whose typical value is found to be close to $300$~m. A natural debate that arises is whether this quantity stems from location uncertainties and errors alone or reveals a genuine structural property of the fault. Its mean value is about $290$~m, which is much larger than the one expected from location errors alone (which is about $30-40$m in the studied catalog). Our conclusion is thus that a fault cannot be considered as an infinitely thin surface, but rather as a very elongated volume with finite thickness. One may think that crossing faults may artificially inflate the thickness of clusters to provide better fits at the crossings. This explanation can be ruled out, as shown by our study of the synthetic example studied in section \ref{appsynthex}, whose correct parameters for the three fault planes were correctly retrieved by our algorithm. It is worth noting that the thickness mentioned above is compatible with the thickness of the damage zone which is commonly observed around faults (Ben Zion and Sammis, 2003). The various faults in the Mount Lewis area may thus not be mature enough to have developed a significant fault core able to localize earthquake ruptures. This last conclusion is compatible with the idea that the Mount Lewis fault is a young structure which is still in its early growth process. Future work will focus on the active faults analyzed by De Joussineau and Aydin in order to check how the segmentation properties at the surface correlates with the size of earthquake clusters at depth.

Fitting the data with an increasing number of kernels allows one to inspect spatial features 
at smaller and smaller scales within the catalog. It is thus akin to a multi-scale analysis. In the latter, the resolution is fixed and one can compute some relevant parameters, such as the number of clusters as a function of scale, for instance. 
The present approach has an inverted logic: choosing the number of clusters, we could infer their position, size and orientation
as well as scaling properties (see expressions (\ref{hyybwqa})).

More work is needed to provide a stronger theoretical formalism of data clustering of multifractal sets of events. Some may see an incompatibilty between the scale invariance of earthquake catalogs and the exponential distribution of the size of clusters, as one would certainly expect a power-law distribution. It is then worth reminding a very popular and deterministic fractal segmentation model, i.e. the Cantor set. In its most classical construction process, surviving segments are divided into three equal parts, one of which being removed (and replaced by a hole) while the two other ones play the role of survivors for the next iteration. At each scale, all surviving segments possess the same size, but the whole set is self-similar. De Joussineau and Aydin analyzed the distribution of fault segments length in various settings and found it to be a log-normal distribution, i.e. this distribution possesses a characteristic size while the faults (or the fault network) seem to display scale-invariant characteristics. Sornette (1991) has 
also provided examples to debunk the misconception that fault length power law distributions are in any way related
to fractal geometrical organization of their network.


{}

\end{article}

\clearpage

\begin{table}
\caption{\label{table_synth} Table of correspondence between the parameters of the input fault planes used to generate the synthetic catalog shown in Figure \ref{synth_0_plane} and those of the fitting kernels found by the EM algorithm. Longitudes and latitudes refer to those of the barycenters of the faults (or kernels). For each fault label, the first line corresponds to the input fault, while the second line corresponds to the fitted parameters that have been determined by the algorithm.}
\begin{center}
\hskip -2.3cm
\begin{tabular}{lccccccccc}
\hline
\\
Label & Long.     & Lat.    & Depth (km) & Strike & Dip  & Length(km) & Width(km) & $\sigma_{3} (km)$ & $N$  \\
\hline \\
A     & -0.100 &  0.000 & 10.000 & 180.00 & 90.00 &  20.000 &  10.000 & 0.006 & 100  \\
      & -0.100 & -0.003 & 10.372 & 180.03 & 89.92 &  19.666 &  10.056 & 0.003 & 100.46 \\
B     & +0.100 &  0.000 & 10.000 & 180.00 & 90.00 &  20.000 &  10.000 & 0.006 & 100  \\
      & +0.100 &  0.006 &  9.791 & 180.05 & 89.96 &  21.241 &  10.014 & 0.003 & 99.58 \\
C     &  0.000 &  0.000 & 10.000 & 270.00 & 90.00 &  40.000 &  10.000 & 0.006 & 200  \\
      & -0.017 &  0.000 & 10.486 & 270.00 & 89.85 &  38.756 &  10.249 & 0.003 & 199.96 \\
\\
\hline
\end{tabular}
\end{center}
\end{table}

\clearpage

\clearpage

\clearpage

\clearpage

\clearpage

\clearpage

\clearpage

\clearpage

\clearpage

\clearpage

\clearpage

\clearpage

\clearpage

\clearpage

\clearpage

\clearpage

\clearpage

\clearpage

\clearpage

\clearpage

\clearpage

\clearpage

\clearpage

\clearpage

\clearpage

\clearpage

\clearpage

\clearpage

\clearpage

\end{document}